\documentclass[twocolumn]{aastex61}

\received{Feb 1, 2019}
\revised{Feb 27, 2019}
\accepted{\today}
\submitjournal{ApJ}
\shorttitle{NGC3079}
\shortauthors{Sebastian et al.}
\begin{document}
\title{The Filamentary Radio Lobes of the Seyfert-Starburst Composite Galaxy NGC\,3079}
\correspondingauthor{B. Sebastian}
\email{biny@ncra.tifr.res.in}
\author[0000-0001-8428-6525]{Biny Sebastian}
\affil{National Centre for Radio Astrophysics (NCRA) - Tata Institute of Fundamental Research (TIFR),\\ S. P. Pune University Campus, Post Bag 3, Ganeshkhind, Pune 411007, India}
\author[0000-0003-3203-1613]{P. Kharb}
\affiliation{National Centre for Radio Astrophysics (NCRA) - Tata Institute of Fundamental Research (TIFR),\\ S. P. Pune University Campus, Post Bag 3, Ganeshkhind, Pune 411007, India}
\author[0000-0001-6421-054X]{C. P. O' Dea}
\affiliation{School of Physics and Astronomy, Rochester Institute of Technology, Rochester, NY 14623, USA}
\affiliation{Physics and Astronomy, University of Manitoba, Winnipeg, Canada}
\author[0000-0000-0000-0000]{E. J. M. Colbert}
\affiliation{Hume Center for National Security and Technology, 900 N. Glebe Rd, Arlington, VA 22203US Army Research Laboratory
Adelphi, USA}
\author[0000-0000-0000-0000]{S. A. Baum}
\affiliation{Carlson Center of Imaging Science, Rochester Institute of Technology, Rochester, NY 14623, USA}
\affiliation{Physics and Astronomy, University of Manitoba, Winnipeg, Canada}

\begin{abstract}

We present results from multi-frequency polarization-sensitive Very Large Array observations of the Seyfert-starburst composite galaxy NGC\,3079. Our sensitive radio observations reveal a plethora of radio ``filaments'' comprising the radio lobes in this galaxy. We analyze the origin of these radio filaments in the context of existing Chandra X-ray and HST emission-line data. We do not find a one-to-one correlation of the radio filaments with the emission line filaments. The north-eastern lobe is highly polarized with polarization fractions $\sim 33\%$ at 5 GHz. The magnetic fields are aligned with the linear extents of the optically-thin filaments, as observed in our, as well as other observations in the literature. Our rotation measure images show evidence for rotation measure inversion in the north-eastern lobe. Our data best fit a model where the cosmic rays follow the magnetic field lines generated as a result of the dynamo mechanism. There could be additional effects like shock acceleration that might also be playing a role.
We speculate that the peculiar radio lobe morphology is a result of an interplay between both the superwinds and the AGN jet that are present in the galaxy. The jet, in fact, might be playing a major role in providing the relativistic electron population that is present in the radio lobes.

\end{abstract}
\keywords{galaxies: Seyfert --- galaxies: jets --- galaxies: individual (NGC\,3079)}

\section{Introduction} 
\label{sec:intro}
Radio observations reveal that most Seyfert galaxies have sub-parsec-scale radio emission in spite of their ``radio-quiet" AGN status \citep{deBruyn76,UlvestadWilson84A,UlvestadWilson84B,Roy94,Thean00,Lal04,Kharb10a,Kharb14a}.  Many of these galaxies with small-scale radio emission show elongated kiloparsec-scale radio structures (KSRs), similar to the radio jets seen in powerful radio galaxies. The kpc-scale radio structures, however, are not always found to be aligned with the small scale jets \citep{Baum93,Colbert96,Gallimore06}. Furthermore, Seyfert radio jets were found to be randomly aligned with respect to the host galaxy major axes \citep{Kinney00,Schmitt01,SchmittKinney02}. Abrupt changes of jet axes were also inferred in several Seyfert galaxies, e.g., NGC\,4151 \citep{Ulvestad98}, NGC\,1068 \citep{Gallimore96}, and Mrk\,6 \citep{Kharb06}. It appears that the radio structures in Seyfert galaxies are more complex than in radio galaxies. In particular, it is unclear if KSRs are purely AGN-driven (either via a jet or an accretion-disk wind) or starburst superwind-driven. It is essential to disentangle these contributions before the truly relevant questions of how radio outflows are produced in low-luminosity AGN (LLAGN) in general, and Seyfert galaxies in particular can be answered. Also, the role played by LLAGN outflows in AGN feedback needs to be examined \citep{croton2006,muller2011,riffel2013}. 

NGC\,3079 is a well known nearby ($z=0.003723$) Seyfert 2 / LINER galaxy with a nuclear starburst, that is a member of an interacting galaxy pair. It is an edge-on spiral galaxy with several regions of star-formation along the disk. Its 8-shaped radio structure with a ring-like feature (``ring'' hereafter) inside its north-east lobe \citep{Duric88,irwin2003,wiegert2015,irwin2019}, its slow-moving ($v\sim0.1c$) parsec-scale radio jet \citep{irwinseaquist1988,baanirwin1995,Middelberg05}, copious amount of hot X-ray gas and emission-line filaments \citep{cecil2001,cecil2002}, have been well documented. NGC\,3079 is a Compton-thick AGN \citep{iyomoto2001} having a disk-like pseudo-bulge.
The origin of the double-lobed morphology observed in the NGC\,3079 is not yet completely understood.
\cite{Duric88} argue that the lobe material is moving at supersonic velocities, though they dismiss the possibility that these are jets which may be precessing based on the difficulty in reproducing the closed loop morphology using these models. They suggest that these are probably winds originating from a nuclear starburst or accretion activity. X-ray emission and emission line imaging shows that NGC\,3079 hosts a superwind \citep{Veilleux94,cecil2001,cecil2002}. Superwinds are believed to be generated when the kinetic energy from supernovae and massive stellar winds are transformed into thermal energy. The heated gas then expands into the lower pressure ambient medium. The direction of the steepest pressure gradient is along the minor axis. The hot gas is easily traced by the X-ray emission, which is a more direct tracer compared to the emission line gas, which is generated at shock fronts as a result of the interaction with the ambient medium. The emission line ratios seen in superwinds are typical of shock-ionized gas as also seen in the case of NGC\,3079 \citep{Veilleux94}. 
Although these authors favored a starburst-wind origin for the radio lobes, they were not able to rule out AGN-driven winds. Several authors favored the jet origin of the lobes \citep{irwinseaquist1988,irwin2003}. \cite{irwin2017} find that NGC\,3079 is one of the few edge-on galaxies in the CHANG-ES sample \citep{irwin2012} which shows radio lobes that stand out in polarized intensity compared to the galactic disk emission which is not as highly polarized.

\cite{2001ApJ...547L.115S, 2015PASJ...67....5M, 2016MNRAS.458.1375L} have studied the molecular constitution and kinematics of NGC\,3079 using various molecular lines in both emission and absorption. Many of these studies present evidence for blue-shifted features or blue-wings which was readily associated with outflowing molecular gas. \cite{2015MNRAS.454.1404S} present evidence for outflowing HI gas. These authors propose that large amounts of cold gas, both molecular and atomic, is being blown out along with the ionized gas. They find that both the AGN and starburst scenarios are powerful enough to drive such outflows. They also point out that such multi-phase gas outflows are being discovered in several AGN driven outflows. The presence of a nuclear starburst in NGC\,3079 was contested by \cite{hawarden95}. They suggest that the molecular gas outflows are driven by the AGN rather than a starburst.

Despite the abundance of literature on NGC\,3079, a consensus has not yet been reached regarding the origin of the kpc sized radio emission. In this paper, we present a polarization-sensitive radio study of NGC\,3079 with the historical and expanded VLA, at multiple radio frequencies and array configurations. We attempt to disentangle the contribution of the starburst, AGN winds and the jet in the formation of double-lobed structure in NGC\,3079. Throughout this paper, we assume $H_0$=73~km~s$^{-1}$Mpc$^{-1}$,  $\Omega_{m}=0.27$ and $\Omega_{vac}=0.73$. At the distance of NGC\,3079, 1$\arcsec$ corresponds to 0.084 kpc. Spectral index $\alpha$ is defined such that flux density at frequency $\nu$ is $S_\nu\propto\nu^\alpha$.

\section{Observations and Data Analysis}
\subsection{VLA Data from 1995}
We observed NGC\,3079 with the VLA at 1.5 and 4.9 GHz with the A, B and C-array configurations in $1995-1996$ (Project ID: AB740). 3C\,286 and 3C\,48 were used as the primary flux density as well as the polarization calibrators, while 1035+564 was used as the phase calibrator for the whole experiment. The data were processed with AIPS using standard imaging and self-calibration procedures. Table~\ref{tab1} lists the observing frequency, the corresponding bandwidth, the VLA configuration, the observation date, the FWHMs of the synthesized beams, the total flux densities of the source at the corresponding resolution and the {\it rms} noise of the final images. AIPS task PCAL was used to solve for the antenna ``leakage'' terms (D-terms) as well as polarization of the calibrators 3C\,286 and 3C\,48. The leakages were typically a few percent. Polarization calibration of the 4.9~GHz A-array data was unsuccessful. This was a consequence of not having acquired enough scans of the polarization calibrator for adequate parallactic angle coverage; there were only two scans of 3C\,286. 

\begin{figure*}
\centerline{
\includegraphics[width=9.0cm,trim=0 0 0 0]{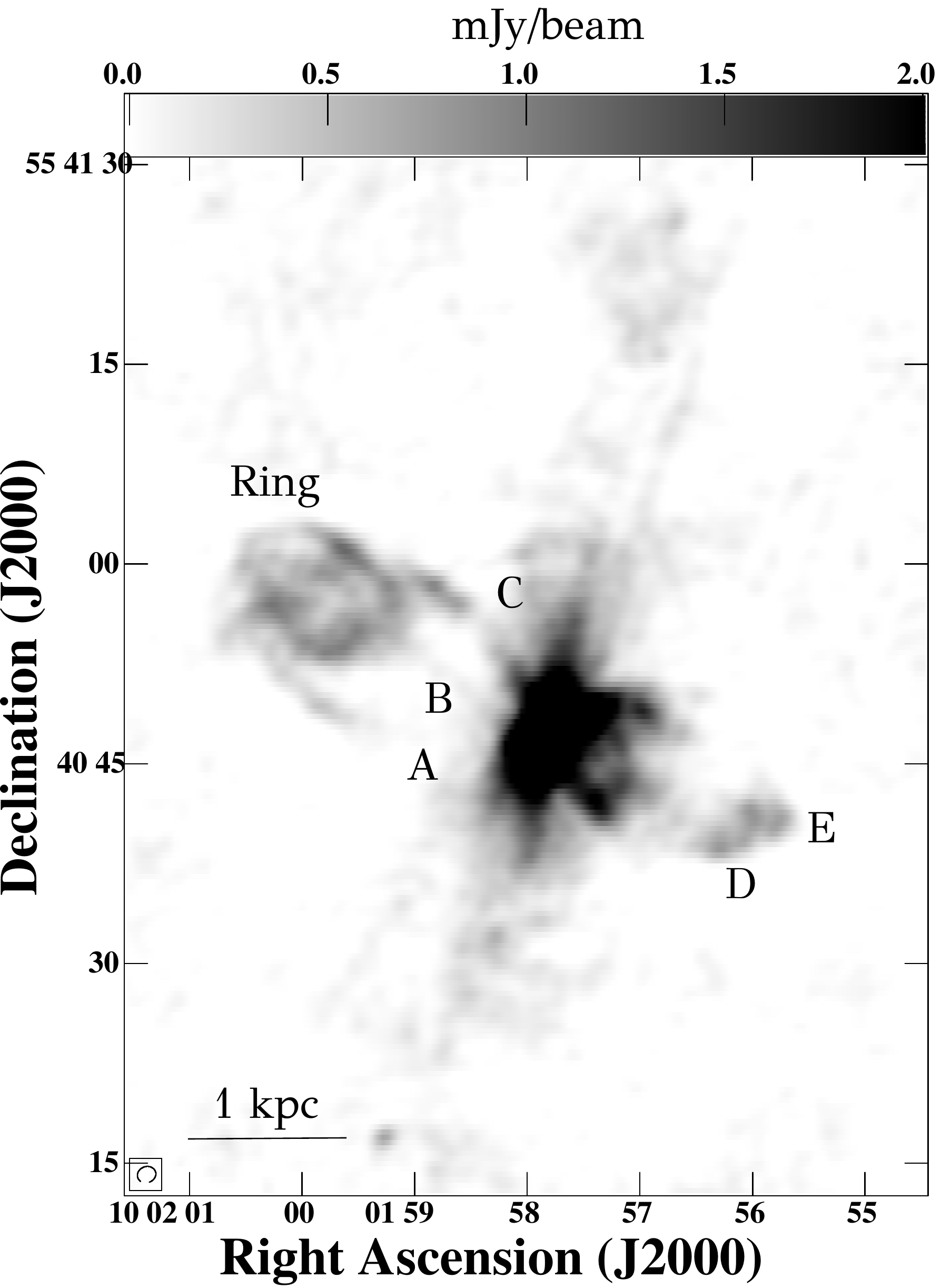}
\includegraphics[width=9.0cm,trim=16 0 0 0]{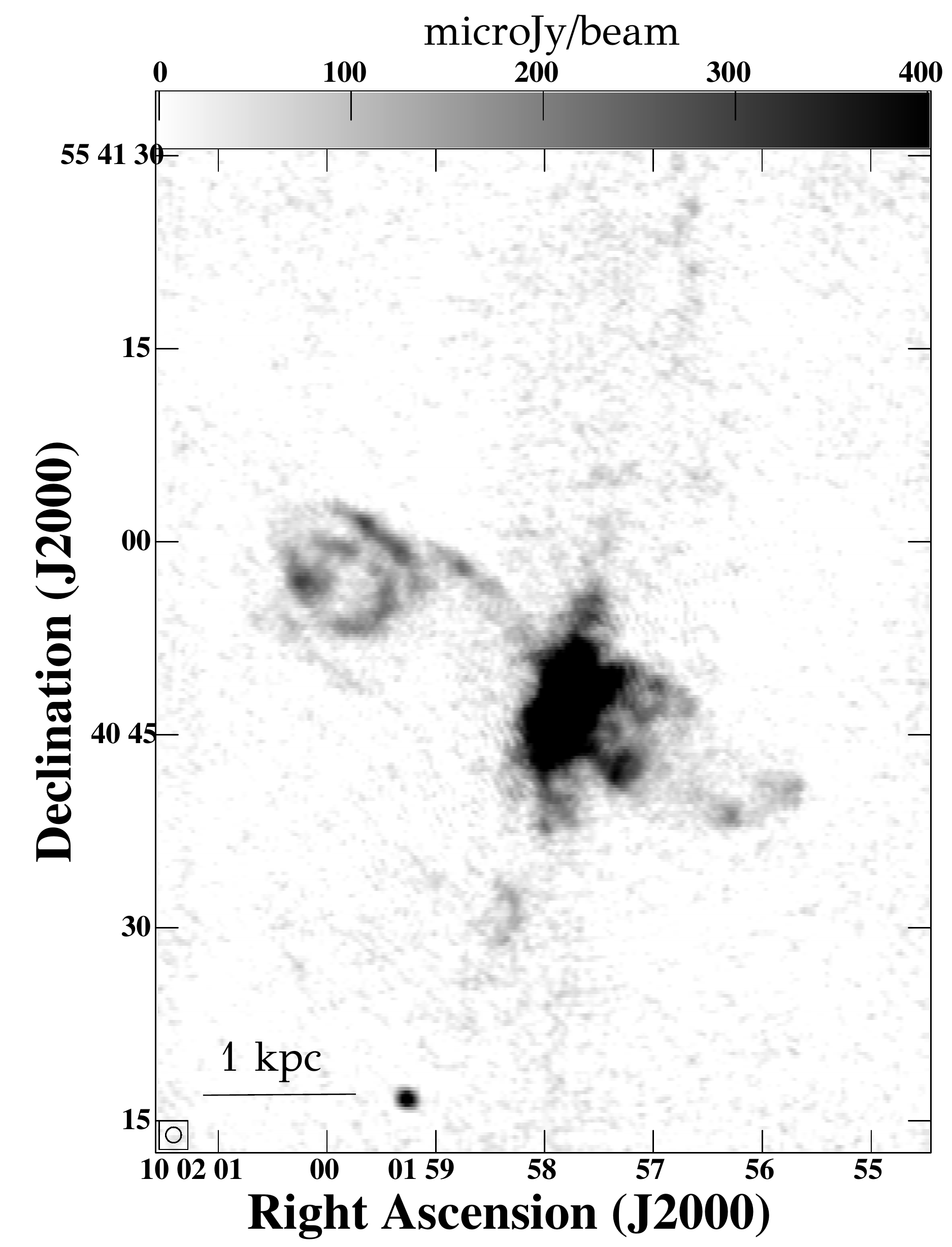}}
\caption{Grey scale images of total intensity showing the filamentary radio structures observed in NGC\,3079 with the (left) VLA A-array at 1.4 GHz with a beam size $1.45\arcsec\times1.19\arcsec$ and P.A., $-80.2\degr$ and (right) VLA B-array at 5 GHz with a beam size, $0.58\arcsec\times0.43\arcsec$ and P.A., 85.8$\degr$.}
\label{fig1}
\end{figure*}

The polarization intensity (PPOL) and polarization angle (PANG) images were created by first running the task IMAGR for Stokes `Q' and `U' and then combining these images using the task COMB. Pixels with intensity values below 3$\sigma$ and angle errors $>10\degr$ were blanked before making PPOL and PANG images, respectively. Fractional polarization images were created using the PPOL and total intensity images where pixels with $>10\%$ errors were blanked. We created two-frequency spectral index images using the AIPS task COMB after convolving images at both frequencies with the same circular beam ($=1.45\arcsec\times1.45\arcsec$). Pixels with total intensity value below 3$\sigma$ were blanked before making the spectral index image. Rotation measure images were created using three frequencies using AIPS tasks MCUBE, TRANS, and RM. For this, we split the two IF data at L-band (IF1 = 1.465 GHz, IF2 = 1.385 GHz) and used the C-band (frequency = 4.860 GHz) data as such. Images were made after first convolving all images with the same circular beam ($=1.45\arcsec\times1.45\arcsec$). Pixels with errors $>13\degr$ were blanked while making the polarization angle images (PANG) shown in Figure~\ref{fig3}. We created depolarization (DP) images by dividing the fractional polarization images at 1.42 GHz and 4.86 GHz using the task 'COMB'. We blanked pixels with fractional errors $>20\degr$ before making the DP image shown in Figure~\ref{figdp}.

\subsection{EVLA Data from 2018}
New polarization-sensitive EVLA data at 5 GHz were acquired in 2018 (project ID: 17B$\_$074; see Table~\ref{tab1}) using the BnA array configuration. 3C286 and 1035+564 were observed as the primary flux + polarization calibrator and phase calibrator respectively. OQ208, which is a strong unpolarized source was observed to calibrate the polarization leakages. Data were reduced using CASA version 5.1.2-4.  Basic calibration and editing of the data were carried out using the CASA pipeline for EVLA data reduction. Polarization calibration was performed on the pre-calibrated data. The model of 3C\,286 was defined using the model parameters which were fed manually to the task `setjy'. The model parameters required for completely defining the source model included the stokes I flux density at the reference frequency, the spectral index, the coefficients of the Taylor expansion of both the degree of linear polarization as a function of the frequency about the reference frequency and the reference frequency itself. The polarization angle of 3C\,286 appears to be a constant across frequencies whereas the degree of polarization varies across frequencies. Therefore we estimated the coefficients by fitting a first-order polynomial using the data from \cite{2013ApJS..206...16P}. The cross-hand delays between both the polarizations were corrected. 
The leakage terms per channel were solved using the CASA task `polcal' with the unpolarized source OQ208. The polarization angle was calibrated using the known polarization angle of 3C286. 

The target NGC\,3079 was then SPLIT out before imaging. We made use of the MT-MFS algorithm in CASA while cleaning to take care of the wideband effects. Imaging and phase-only self-calibration were performed iteratively for three times before a final round of A\&P self-calibration and imaging was performed. The stokes `Q' and 'U' images were made using the final calibrated UV-data file. The EVLA data had 16 spectral windows spanning a bandwidth of 2 GHz. The data set was then divided into four chunks each consisting of 4 spectral windows. The `QU' stokes image cubes of these data sets were then created using task `tclean' in CASA.  We have corrected the polarized emission for Ricean bias using the AIPS task COMB. Finally, the task `rmfit' was used to make an in-band rotation measure image from these cube images. The RM image obtained is shown in \ref{fig5}. The pixels where the errors are greater than 200 rad~m$^{-2}$ were blanked. The RM image is shown in the left panel of figure~\ref{fig5}.

\begin{table*}
\begin{center}
\caption{Observation log and observed parameters.}
\begin{tabular}{cccccccc} \hline \hline
Frequency&Bandwidth& VLA Array &Observation & Beam, PA &Total Flux Density&r.m.s. \\
(GHz)&(MHz)& Configuration& Date&(arcsec$^2$, $\degr$)&(mJy)&($\mu$Jy beam$^{-1}$) \\
\hline
4.86&50&A&07/21/1995&0.45$\times$0.31, 81.1 & 99.4&27\\
4.86&50&B&11/02/1995&0.58$\times$0.43, 85.8 & 161.0&16 \\
1.42&50&A&07/03/1995&1.45$\times$1.19, $-80.2$ & 509.0&53 \\
1.42&50&C&02/17/1996&14.82$\times$13.03, 62.2 & 838.0&80 \\
5.50&2048&BnA&02/16/2018&1.10$\times$0.47, $-87.08$ &260.3& 8.6 \\
\hline
\label{tab1}
\end{tabular}

\end{center}
\end{table*}

\begin{figure*}
\centerline{
\includegraphics[width=10cm,trim=60 80 0 60]{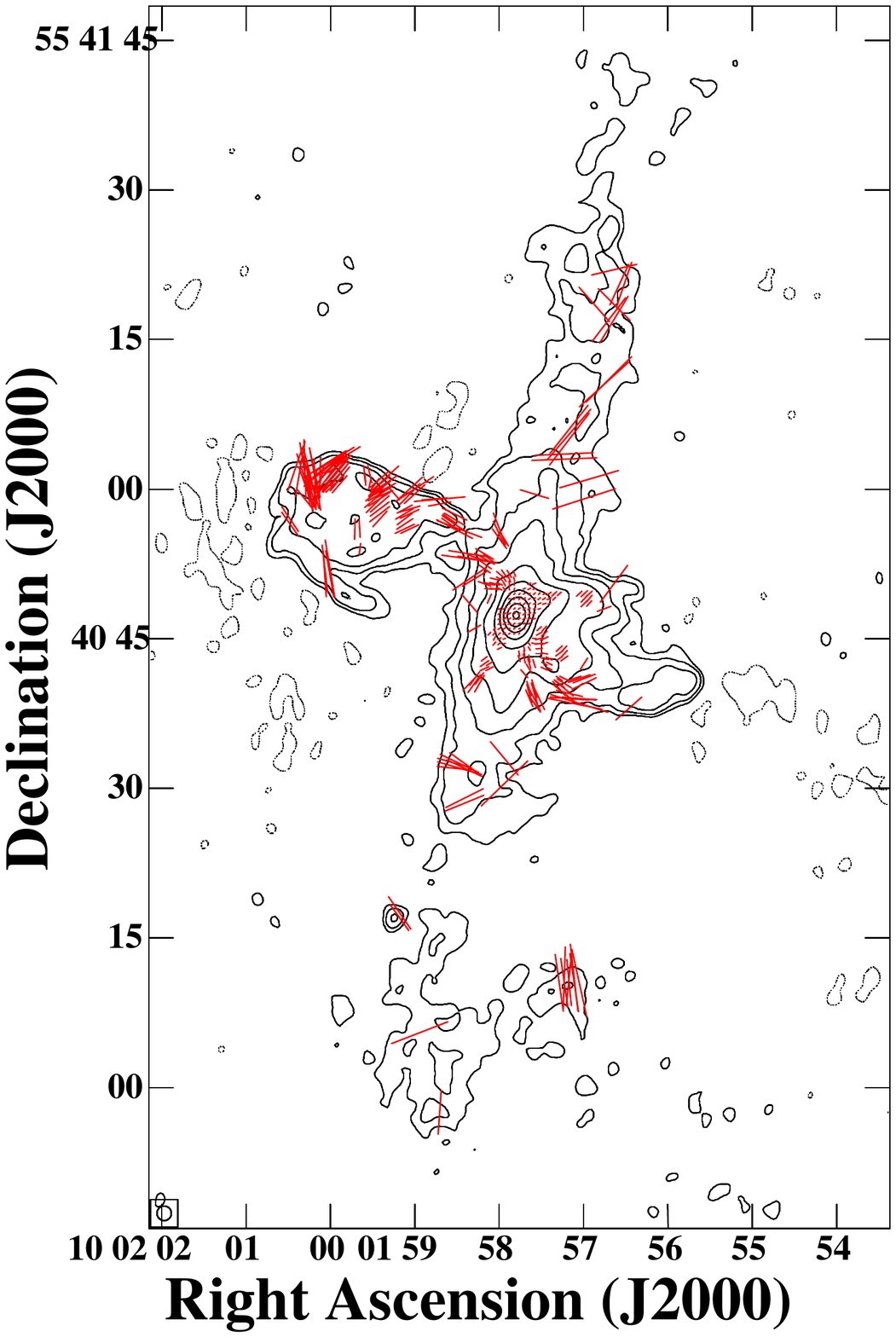}
\includegraphics[width=8.8cm,trim=110 180 0 80]{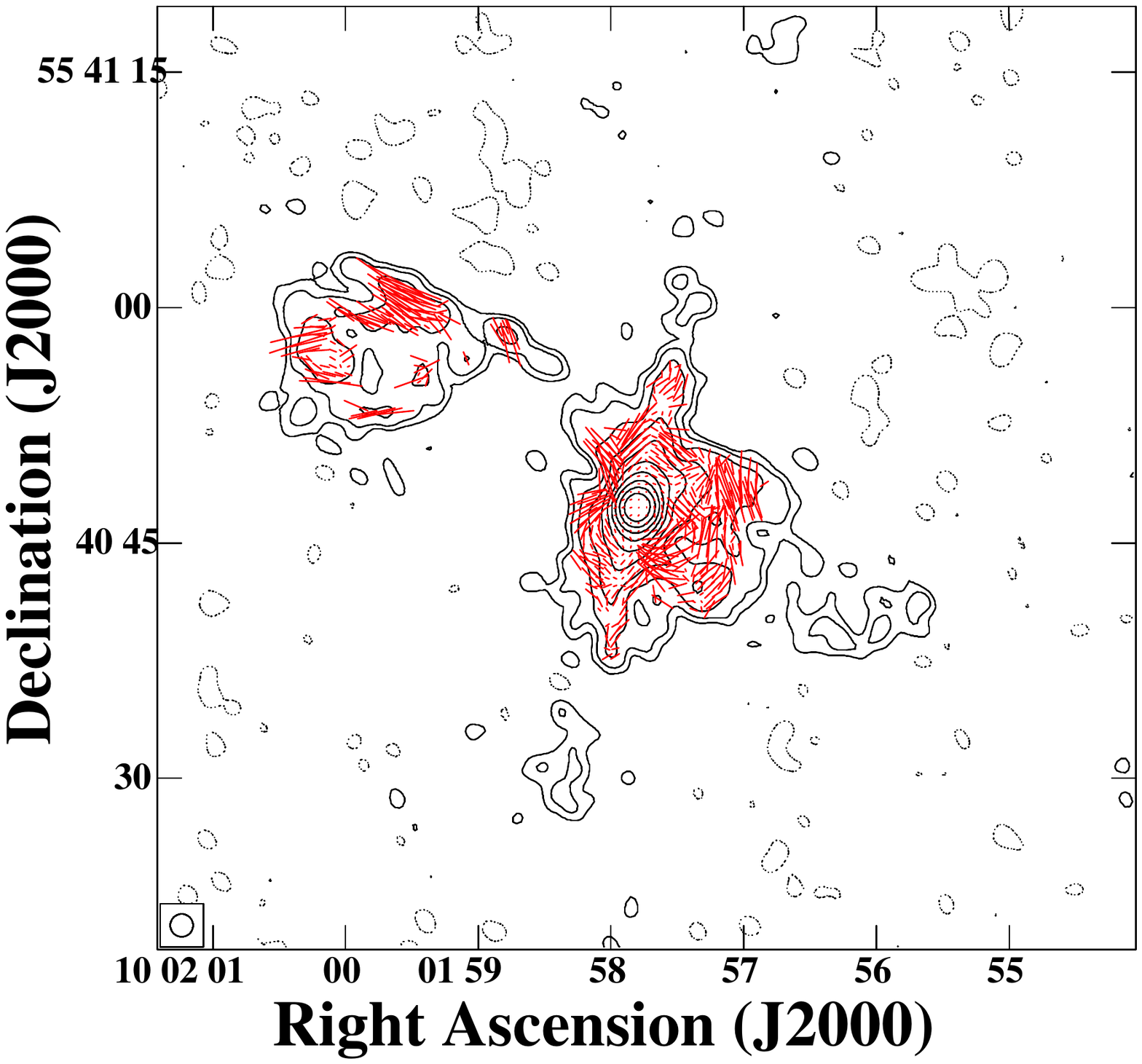}}
\caption{(Left) 1.4 GHz VLA A-array image with contour levels 0.850 $\times$ (-0.170, 0.170, 0.350, 0.700, 1.400, 2.800, 5.600, 11.25, 22.50, 45, 90) mJy beam$^{-1}$
 and (right) 5 GHz VLA B-array image with contour levels 0.800 $\times$ (-0.085, 0.085, 0.170, 0.350, 0.700, 1.400, 2.800, 5.600, 11.25, 22.50, 45, 90) mJy beam$^{-1}$ of NGC\,3079 with magnetic field vectors whose length is proportional to the fractional polarization superimposed in red.}
\label{fig2}
\end{figure*}

\section{Radio Morphology: Lobes, Ring, Galaxy}
\label{secmorph}
\begin{figure*}
\centerline{
\includegraphics[width=18cm,trim=15 0 0 20]{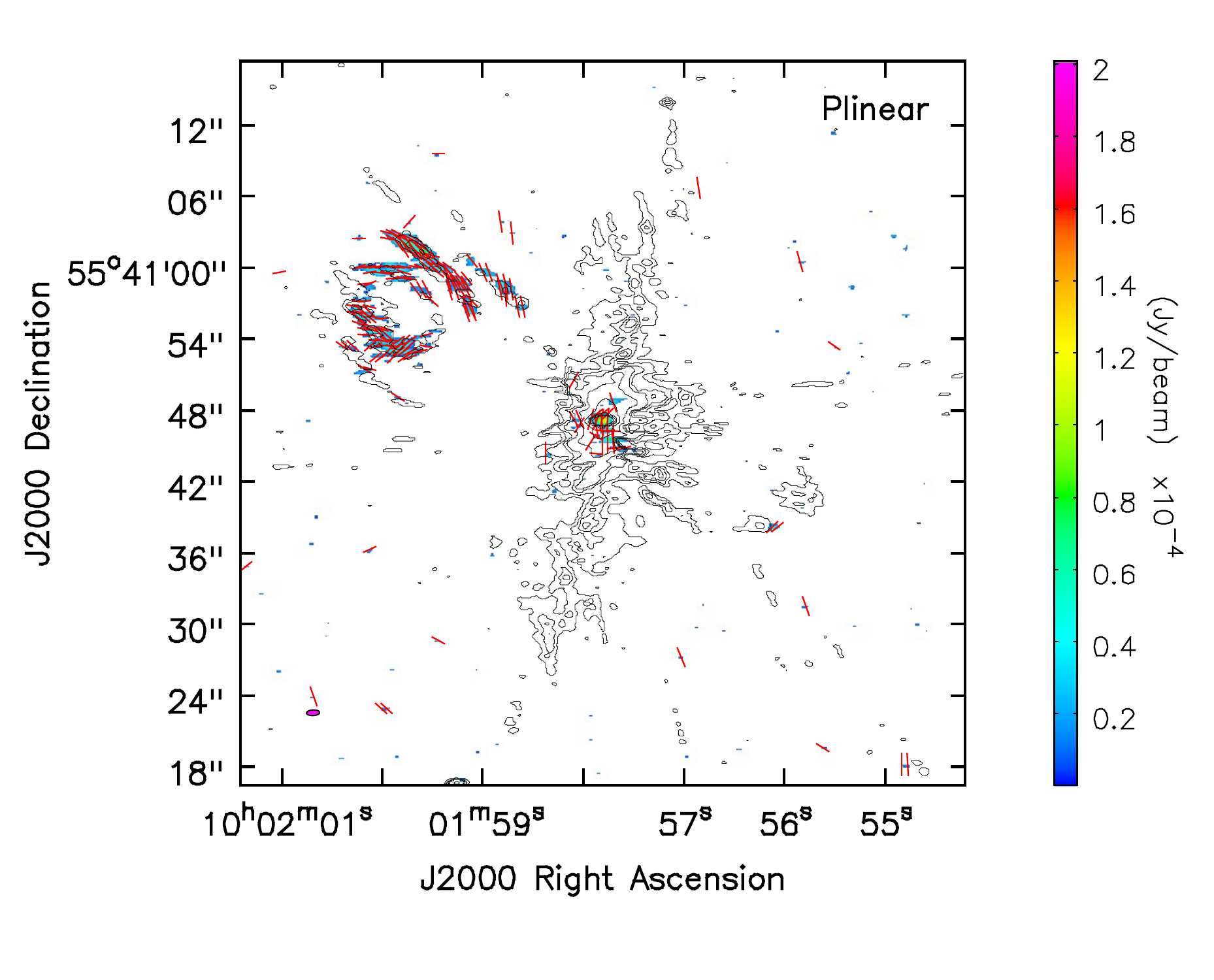}}
\caption{New EVLA 5 GHz image of NGC\,3079 with total intensity in contours and polarized intensity in colour. The magnetic field vectors are shown in red}. Contour levels represent 25.5$\times$(2, 4, 6, 8, 16, 64, 256, 512, 1024) $\mu$Jy beam$^{-1}$. The beam shown in the bottom left corner as a filled magenta ellipse is of size $1.10\arcsec \times 0.47\arcsec$ and P.A. is $-87\degr$.
\label{fig3}
\end{figure*}

The total intensity and linearly polarized emission images for NGC\,3079 are presented in Figures \ref{fig1} to \ref{fig3}. The 1.4~GHz image in Figure \ref{fig1} (right panel) shows the north-eastern lobe with an extent of 27$\arcsec$ (=2.3 kpc) and the south-western lobe with an extent of 20$\arcsec$ (=1.7~kpc) in exquisite detail, especially the extremely filamentary nature of the lobes. The north-eastern lobes show three distinct filaments annotated as A, B and C two of them forming the ``edges" of the radio bubble/lobe. The filament at the southern edge (A) extends up to the top edge of the lobe. This filament shows increased brightness towards the top edge of the lobe. The northern filament (C) can also be traced till the lower end of the ``ring-like" structure after which the trajectory is not clear. This filament also appears brighter, farther away from the core. 
It is probable that the ``ring" is a 3D-shell-like structure seen in projection.
The ``ring-like" feature at the top of the north-eastern lobe appears to be comprised of additional radio filaments. They are the brightest filaments and do not appear to be connected to the center of the galaxy. 

The south-western lobe, on the contrary, is not comprised of as many filamentary structures.
Yet, the edges of the lobe, especially at the base, appear brighter compared to the emission from the inner parts. Further away, towards the southernmost tip, there is an increase in brightness where two filaments, namely E and D, undergo bending and point towards the nuclear core. Unlike the north-eastern lobes, there are no standalone filamentary structures apparently disconnected from the core.

\subsection{Host Galaxy Emission}
\label{hostgal}
\begin{figure*}
\centerline{
\includegraphics[width=7cm,trim=0 0 0 0]{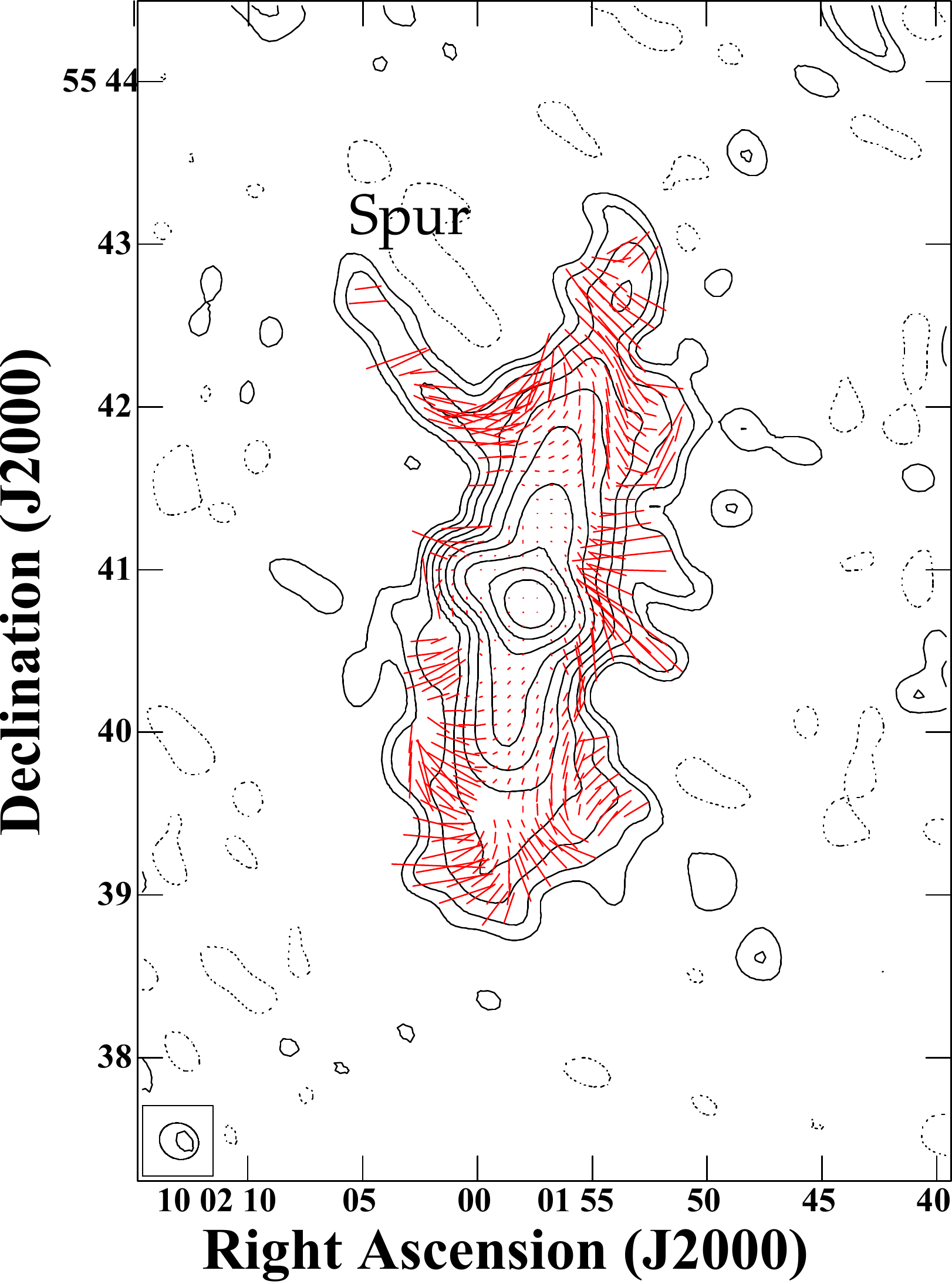}
\includegraphics[width=8.2cm,trim=30 80 20 120]{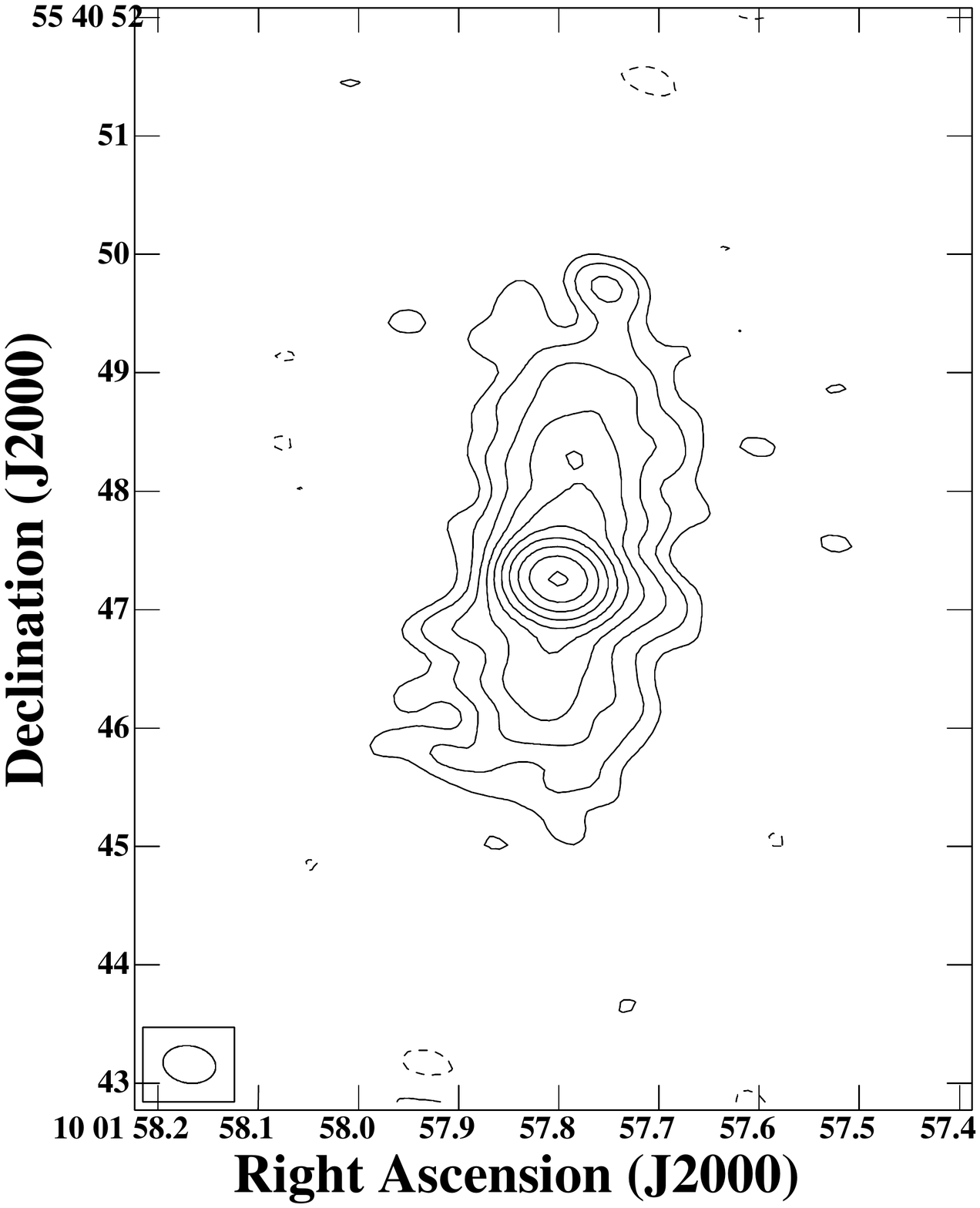}}
\caption{(Left) 1.4 GHz VLA C-array image of NGC\,3079 with contour levels = $3.0 \times$ (-0.085, 0.085, 0.170, 0.350,
0.700, 1.400, 2.800, 5.600, 11.25, 22.50, 45, 90)~mJy~beam$^{-1}$ and (right) 5 GHz VLA A-array image with contour levels = 0.61 $\times$ (-0.085, 0.085, 0.170, 0.350,0.700, 1.400, 2.800, 5.600, 11.25, 22.50, 45, 90)~mJy~beam$^{-1}$. The magnetic field vectors with length proportional to fractional polarization are superimposed in red.}
\label{fig6}
\end{figure*}

Figure \ref{fig6} shows a higher resolution image (VLA A-array  5~GHz) and a lower resolution image (VLA C array 1.4~GHz). The lower resolution image reveals the large extent of host galaxy radio emission. The lobes are seen embedded slightly above the disk in the low-resolution view. Apart from the lobes, there are several features which protrude out of the disk. A prominent spur-like feature is seen rising up to $\sim$8 kpc above the disk at nearly 4 kpc in the north-east direction along the disk from the center. It is also interesting to note that most of these protrusions are highly polarized. Such spurs were previously seen in other starburst systems hosting a radio halo like NGC\,253 \citep{carilli1992}. They suggest that the material from the disk gets convected up and is driven by both cosmic ray pressure as well as thermal pressure. The protrusion can also be explained as galactic fountains \citep{shapiro1976,bregman1980}.

The 615~MHz GMRT image by \cite{irwin2003} shows an entire loop-like structure coming out of the disk plane of the galaxy. The `spur' appears spatially coincident with the eastern-most edge of the loop. At the lower frequency, the halo comprises of much more extended structures than seen in our own images, whereas the `ring' is seen clearly even in our higher frequency maps.
The higher resolution image does not show any lobe emission whereas the emission from the disk which is aligned very well with the outer disk is very prominent. 

\subsection{Spectral and Polarization properties}
\label{spixpolprop}
\begin{figure}
\centerline{
\includegraphics[width=11cm,trim=50 0 0 50]{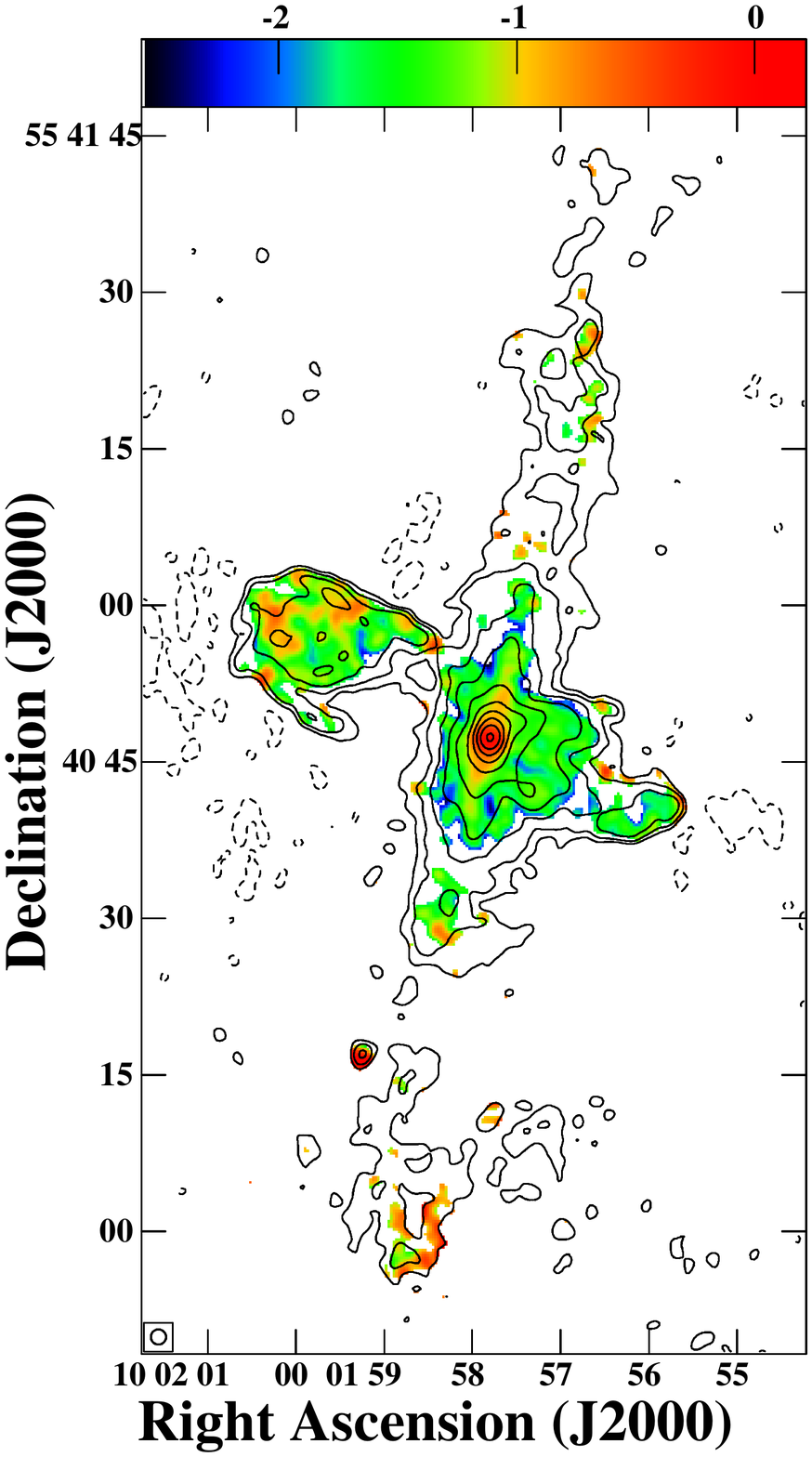}}
\caption{The 1.4 - 5 GHz VLA spectral index image of NGC\,3079 in color superimposed by the 1.4~GHz radio contours. The circular beam shown in the left corner is of size $1.45\arcsec \times 1.45\arcsec$.} 
\label{fig4}
\end{figure}

The 1.4 GHz - 5 GHz spectral index image of NGC\,3079 is shown in Figure~\ref{fig4}. The spectrum is mostly steep (mean $\alpha \sim -1.2\pm 0.13$) except at the core. 
The spectral index image shows that the inner parts of the disk have a flatter spectrum compared to the rest of the source. The core possesses a flat spectrum (alpha $\sim$ $-0.3\pm0.002$) pointing towards current AGN activity as is substantiated by the presence of a relativistically moving VLBI jet \citep{irwinseaquist1988,trotter1998,sawada2000,middelberg2007}. The spectral index image also reveals regions of flat spectrum along the edges of the ring. The spectral index becomes as flat as -0.5 in certain locations along the edges.

The fractional polarization ($FP=P/I$) images reveal the ``ring'' to have $FP = 16\pm5\%$ at L-band and $FP = 33\pm9\%$ at C-band. For the central part of the galaxy showing polarization, we find $FP = 4\pm 1\%$ at L-band and $FP = 16\pm3\%$ at C-band. For the L-band C-array image with a resolution of $\sim14\arcsec$ where the KSR is not clearly resolved (Figure~\ref{fig6}), we find that the average fractional polarization is $FP = 19 \pm 3\%$. Interestingly, the ``spur" like feature to the north-east of the host galaxy disk is very highly polarized:  $FP = 40 \pm 7\%$. Images made with fractional polarization pixels blanked for errors greater than 20\% suggest that at L-band, the ``ring" and the galaxy have $FP=14\pm4$\% and $FP=4\pm1$\%, respectively, and at C-band, the ``ring" and the galaxy have $FP=40\pm12$\% and $FP=24\pm7$\%, respectively. 

The new EVLA image at 5 GHz mostly detects the ring-like feature (Figure~\ref{fig3}); the fractional polarization is $37\pm8\%$ in the ``ring" and $6\pm2\%$ in the galaxy. 
The filaments comprising the ``ring'' are more clearly delineated in the new image in polarized light.
Our in-band RM image indicates that the average RM values are in the range of $\pm50-100$~rad~m$^{-2}$, similar to what was previously observed in the ``ring" by \citep{cecil2001}. 

\begin{figure*}
\centering{
\includegraphics[width=8.8cm,trim=60 180 0 90]{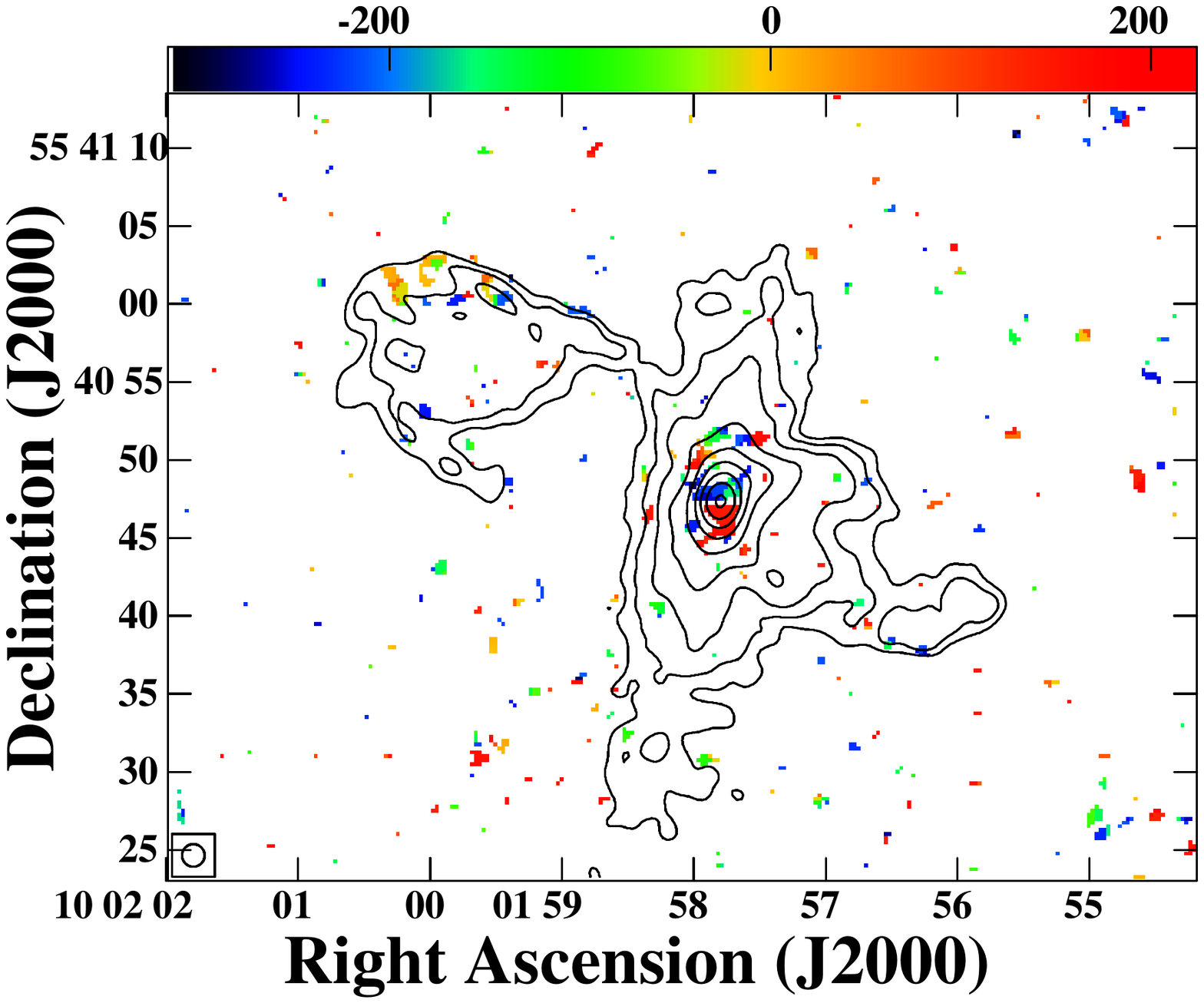}
\includegraphics[width=8.8cm,trim=30 330 30 0]{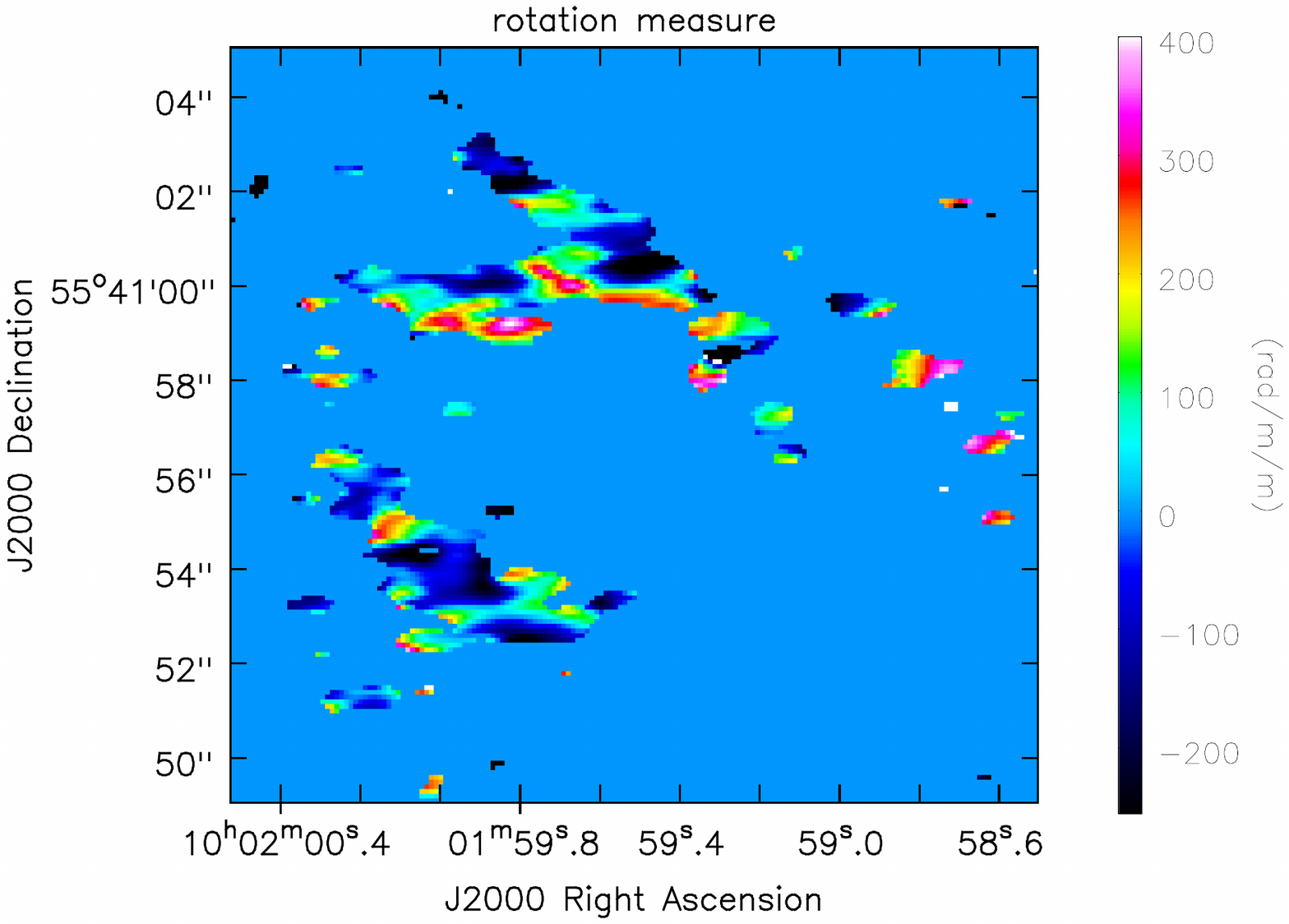}}
\caption{(Left) The three-frequency VLA rotation measure image in color superimposed by the 1.4~GHz radio contours and (right) the color in-band EVLA rotation measure image of the north-eastern lobe. The circular beam shown in the left corner of the left panel is of size $1.45\arcsec \times 1.45\arcsec$. }
\label{fig5}
\end{figure*}

\begin{figure}
\centerline{
\includegraphics[width=7cm,trim= 100 100 60 150]{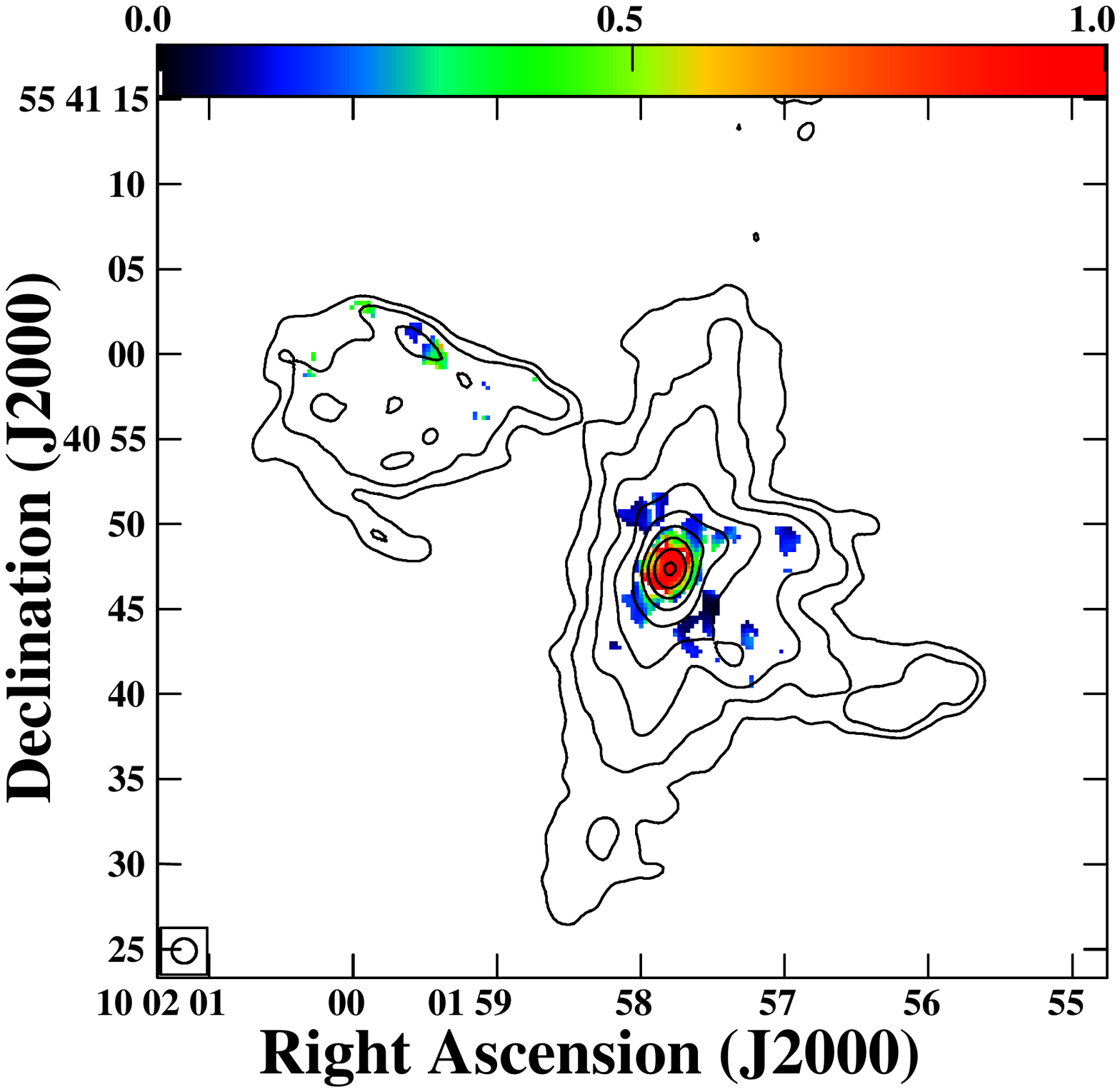}}
\caption{The VLA depolarization image of NGC\,3079 in color made using the 1.4 and 5~GHz image, superimposed by the 1.4~GHz radio contours. The circular beam shown in the left corner is of size $1.45\arcsec \times 1.45\arcsec$.}
\label{figdp}
\end{figure}

The three-frequency RM image (see figure~\ref{fig5} left panel) reveals an RM = +150 rad~m$^{-2}$ with an error of 30 rad~m$^{-2}$, for the southern part of the host galaxy emission. For the western jet region, RM = $-210$~rad~m$^{-2}$ with an error of 15~rad~m$^{-2}$. We have also estimated the depolarization parameter, DP, as DP=$m_l/m_h$, where $m_l$ and $m_h$ are the fractional polarization at low and high frequencies, respectively. The DP image is shown in Figure~\ref{figdp}. A higher value of DP implies lower depolarization. We find that the average DP = 0.3 in the ``ring", and 0.4 in the galaxy; the average DP approaches 0.8 in the central regions of the galaxy. The higher depolarization around the ring could be suggestive of greater confinement.

\section{Energetics of Radio lobes}
\label{globprop}

NGC\,3079 has a flared and clumpy accretion disk \citep{jian2008} with detections of water maser emission on milliarcsec scales \citep{kondratko2000}. 
\cite{kondratko2000} have estimated a black hole mass of 2$\times10^6 M_{\sun}$ in NGC\,3079 using a VLBA study of water masers within a radius of 0.4 pc. \cite{iyomoto2001} find that X-ray emission is heavily obscured and the luminosity after correction for absorption turns out to $10^{42}-10^{43}$~ergs~s$^{-1}$. 
This luminosity would imply that the source is accreting at a rate of 0.01 to 0.1 times its Eddington luminosity. 
The accretion rates are consistent with that seen in Seyfert galaxies \citep{ho2008}.

The parameters derived from our radio observations of NGC\,3079 are listed in Table~\ref{tab3}. 
The parameters derived were total radio luminosity ($L_{rad}$), magnetic field at minimum pressure ($B_{min}$), minimum energy ($E_{min}$), minimum pressure ($P_{min}$), total energy ($E_{tot}$), energy per unit volume for a given filling factor or energy density ($E_{den}$) and electron lifetime from synchrotron and inverse Compton (over CMB) radiative losses ($t_{life}$). 
The minimum pressure parameters were estimated for the north-east lobes, south-west lobes, and the galactic disk separately using equations (1)-(5) from \cite{odea1987}. The lifetimes were estimated using the relation given in \cite{perez2007}.

The B-array image at 5 GHz and the A-array image at 1.4 GHz were used  while estimating the flux densities as these images had matched baseline coverages. These images were then convolved with the same beam size.
Rectangular regions were used for estimating the flux densities.  
The radio spectrum was assumed to extend from 10 MHz to 10 GHz. The proton to electron energy ratio was assumed to be unity. However, \citet{beck2005} have noted that protons are more dominant than electrons according to currently proposed mechanisms for the origin of cosmic rays. In this scenario, our magnetic field values will be underestimated by a factor of $(K_0 +1)^{\frac{1}{\alpha+3}}$, where $K_0$ is the ratio of proton to electron number densities. Hence, for typical values of spectral indices seen in the lobes of NGC\,3079 ($\alpha \sim -$1) and assuming $K_0$=100, the magnetic fields are underestimated by a factor of 3. However, in the lobes of radio galaxies, the classical equipartition magnetic fields often match with those measured from X-rays \citep{brunetti1997}, justifying our choice of K$_0$. Table~\ref{tab3} lists the parameters corresponding to a volume filling factor ($\phi$) values of 1, 0.5 and $10^{-4}$ \citep{blustin2009}. 

\begin{figure*}
\centering{
\includegraphics[width=8.8cm, trim = 30 20 0 0]{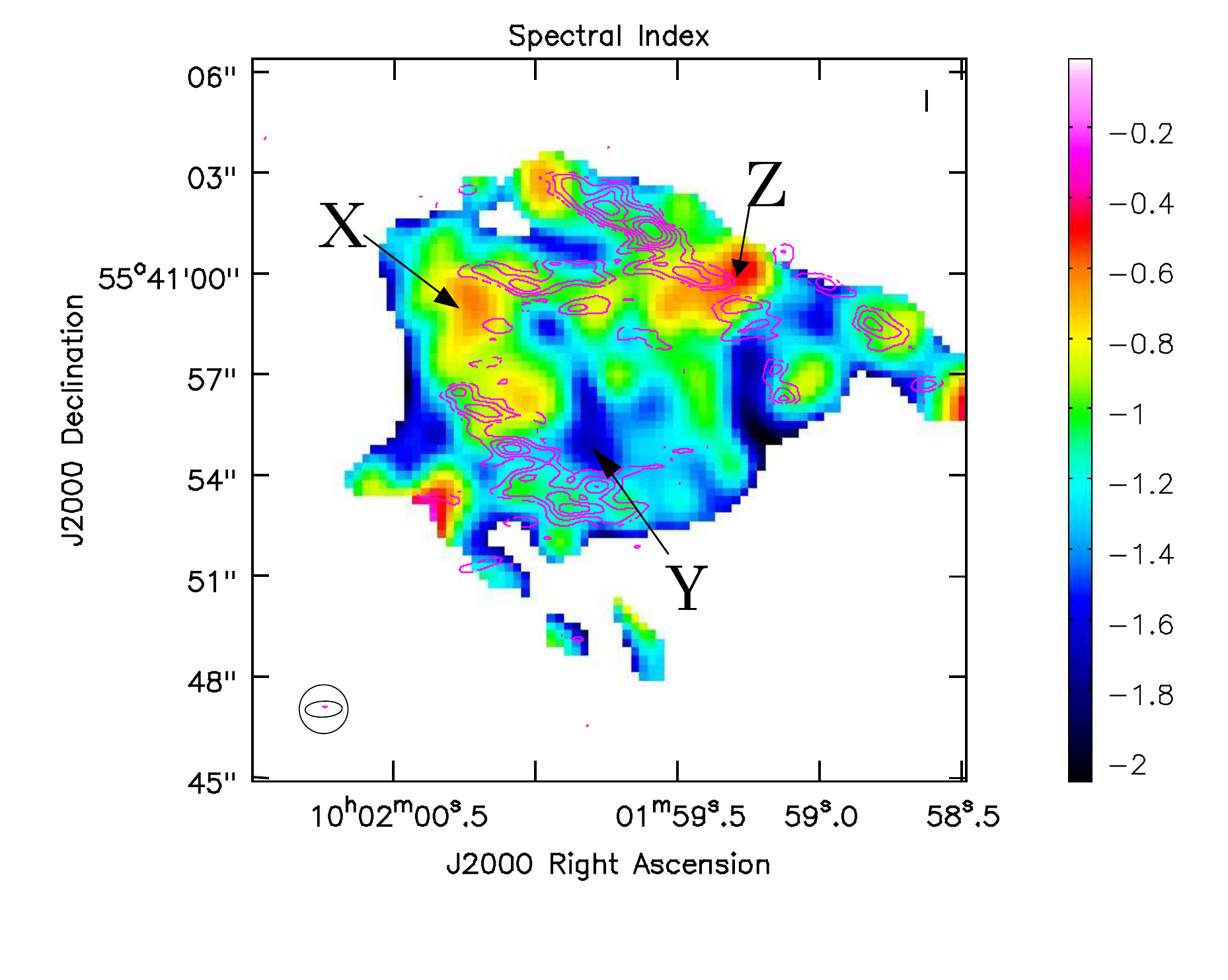}
\includegraphics[width=8.8cm, trim = 30 20 0 0]{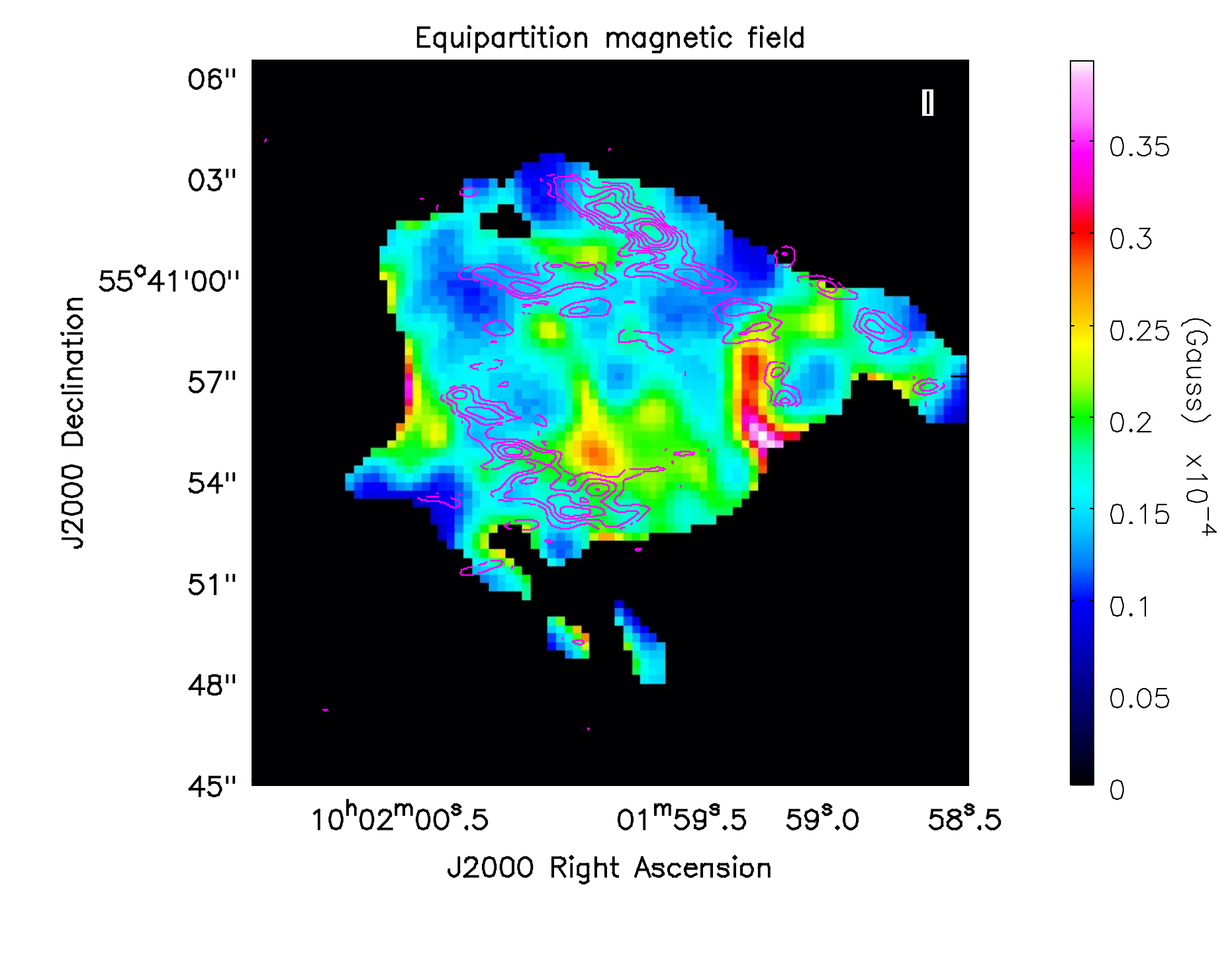}}
\caption{(Left) The spectral index image and (right) equipartition magnetic field image in colour, of the north-eastern lobe made using the 1.4~GHz and 5~GHz VLA images. The pink contours denote the EVLA polarized intensity at 5~GHz. The left bottom corner shows the 1.4 and 5~GHz VLA circular beams enclosing the elliptical EVLA beam at 5~GHz.}
\label{figmagdrap}
\end{figure*}

To further investigate the underlying mechanism we derive the equipartition magnetic fields for every pixel in a similar manner. A spatially resolved image of the spectral index was made instead of assuming an average spectral index value. A volume filling factor equal to unity was used.  A cylindrical volume with a height equal to the width of the lobe (0.8 kpc) was assumed. The radius of the cylinder was chosen to be the same as that of the synthesized beam of the image. The pixels with flux density less than 3$\sigma$ were blanked. The spectral index image and the magnetic field image of the source are shown in Figure~\ref{figmagdrap}. The EVLA polarized intensity image at 5 GHz is shown overlaid in contours. 
The spectral index image shows flatter spectral indices along the ring. 

The spectral index and the magnetic field structure is quite complex. Several regions along the ring-like those marked X \& Z in Figure~\ref{figmagdrap}, shows flat spectrum, whereas region Y show steep spectrum. Region Y forms the inner part of the ring as seen at 1.4 GHz. The width of the ring is not as large at 5 GHz, where most of the emission is confined to the outer edges where the spectral indices are flatter as well. 
Region Y also shows high magnetic field values though polarized emission is not detected at 5 GHz probably because of the steep spectrum which reduces the total intensity and thereby polarized flux density or because of the absence of ordered magnetic fields or both.

\begin{table*}[h]
\begin{center}
\caption{Source parameters estimated from the ``minimum energy'' condition}
\begin{tabular}{cccccccc} \hline \hline
Component &$L_{rad}$  &$B_{min}$& $E_{min}$ & $P_{min}$ & $E_{tot}$  & $E_{den}$ &$t_{life}$  \\
          &(erg/s)  &(Gauss)  & (ergs)& (dynes/cm$^2$)&  (ergs) & (ergs/cm$^3$) & (Myrs) \\
\hline
\hline 
filling factor = 1.0 \\
\hline 

NE lobes &3.0$\times 10^{38}$ & 3.88$\times 10^{-5}$ & 1.49$\times 10^{55}$ & 1.40$\times 10^{-10}$ & 1.86$\times 10^{55}$ & 3.0$\times 10^{-10}$ & 2.74 \\
SW lobes &5.0$\times 10^{38}$ & 4.75$\times 10^{-5}$ & 1.63$\times 10^{55}$ & 2.10$\times 10^{-10}$ & 2.03$\times 10^{55}$ & 4.5$\times 10^{-10}$ & 2.03 \\

Disk &5.0$\times 10^{38} $ & 3.59$\times 10^{-5} $ & 2.52$\times 10^{55} $ & 1.2$\times 10^{-10} $ & 3.15$\times 10^{55} $ & 2.6$\times 10^{-10} $ & 3.07 \\

filling factor = 0.5 \\
\hline 0

NE lobes &3.0$\times 10^{38}$ & 4.73$\times 10^{-5}$ & 1.10$\times 10^{55}$ & 2.08$\times 10^{-10}$ & 1.38$\times 10^{55}$ & 4.5$\times 10^{-10}$ & 2.04 \\
SW lobes &5.0$\times 10^{38}$ & 5.79$\times 10^{-5}$ & 1.21$\times 10^{55}$ & 3.11$\times 10^{-10}$ & 1.51$\times 10^{55}$ & 6.7$\times 10^{-10}$ & 1.51 \\

Disk &5.0$\times 10^{38} $ & 4.38$\times 10^{-5} $ & 1.87$\times 10^{55} $ & 1.78$\times 10^{-10} $ & 2.34$\times 10^{55} $ & 3.8$\times 10^{-10} $ & 2.29 \\

\hline 
filling factor =1.0$\times 10^{-4}$\\
\hline 

NE lobes &3.0$\times 10^{38}$ & 0.00054 & 3.00$\times 10^{53}$ & 2.70$\times 10^{-8}$ & 4.00$\times 10^{53}$ & 5.8$\times 10^{-8}$ & 0.05 \\
SW lobes &5.0$\times 10^{38}$ & 0.00066 & 3.00$\times 10^{53}$ & 4.05$\times 10^{-8}$ & 4.00$\times 10^{53}$ & 8.7$\times 10^{-8}$ & 0.04 \\
Disk &5.0$\times 10^{38} $ & 0.0005 & 5.00$\times 10^{53} $ & 2.30$\times 10^{-8} $ & 6.00$\times 10^{53} $ & 5.0$\times 10^{-8} $ & 0.06 \\

\hline

\hline
\label{tab3}
\end{tabular}

{Column 1: Radio source components including the north-east (NE), southwest (SW) lobes and the galactic ``Disk'' minus the AGN. Column 2: Total radio luminosity. Columns 3, 4, 5 : ``Mininum energy'' magnetic field strength,  energy and pressure, respectively. Column 6: Total energy. Column 7: Total energy density. Column 8: Electron radiative lifetimes via synchrotron and inverse Compton (over CMB) losses. See Section~\ref{globprop}} 

\end{center}
\end{table*}

\label{magfield}

\section{The origin of the ring and filamentary morphology}
\label{orig}

In this section, we try to understand the origin of the peculiar morphology of the radio lobes in NGC\,3079 using our radio observations in conjugation with the existing results from the X-ray and emission line data from the literature.

\cite{cecil2002} find evidence for a large scale diffuse X-ray halo surrounding the radio superbubble which is delineated by $H\alpha$ filaments at wide angles. These $H\alpha$ filaments were explained as the contact discontinuity between the shocked wind and the galaxy halo gas. The inner four filaments of $H\alpha$ emission form the walls of the superbubble \citep{cecil2001}. \cite{cecil2001} infer that the radial velocities in these filaments increase with distance from the disk.  They also find that the four filaments get disrupted at some height which was explained as a result of Rayleigh-Taylor instabilities that may develop in the case of wind-driven superbubbles.

\begin{figure*}
\centerline{
\includegraphics[width=8cm, trim=35 0 0 0 ]{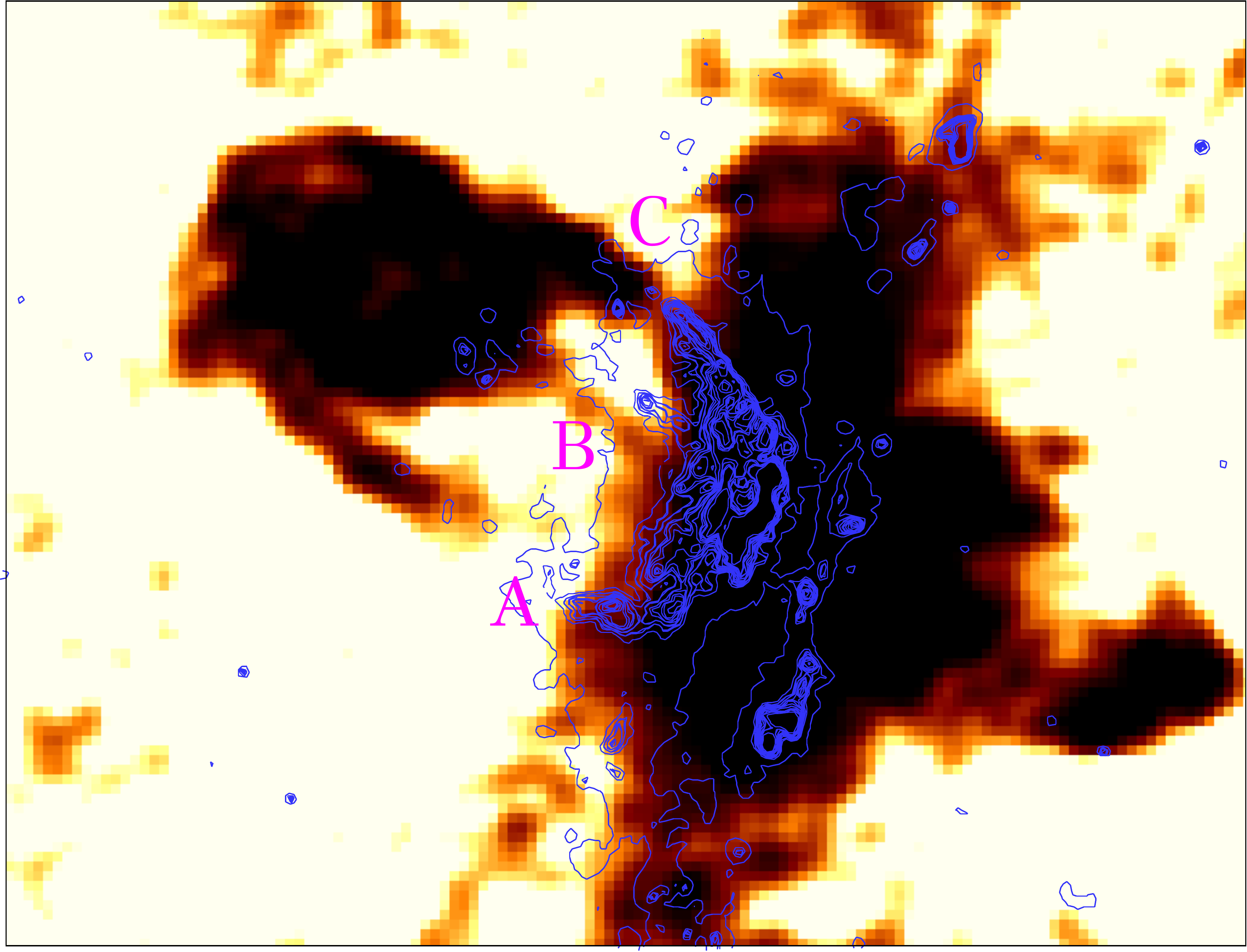}
\includegraphics[width=8cm]{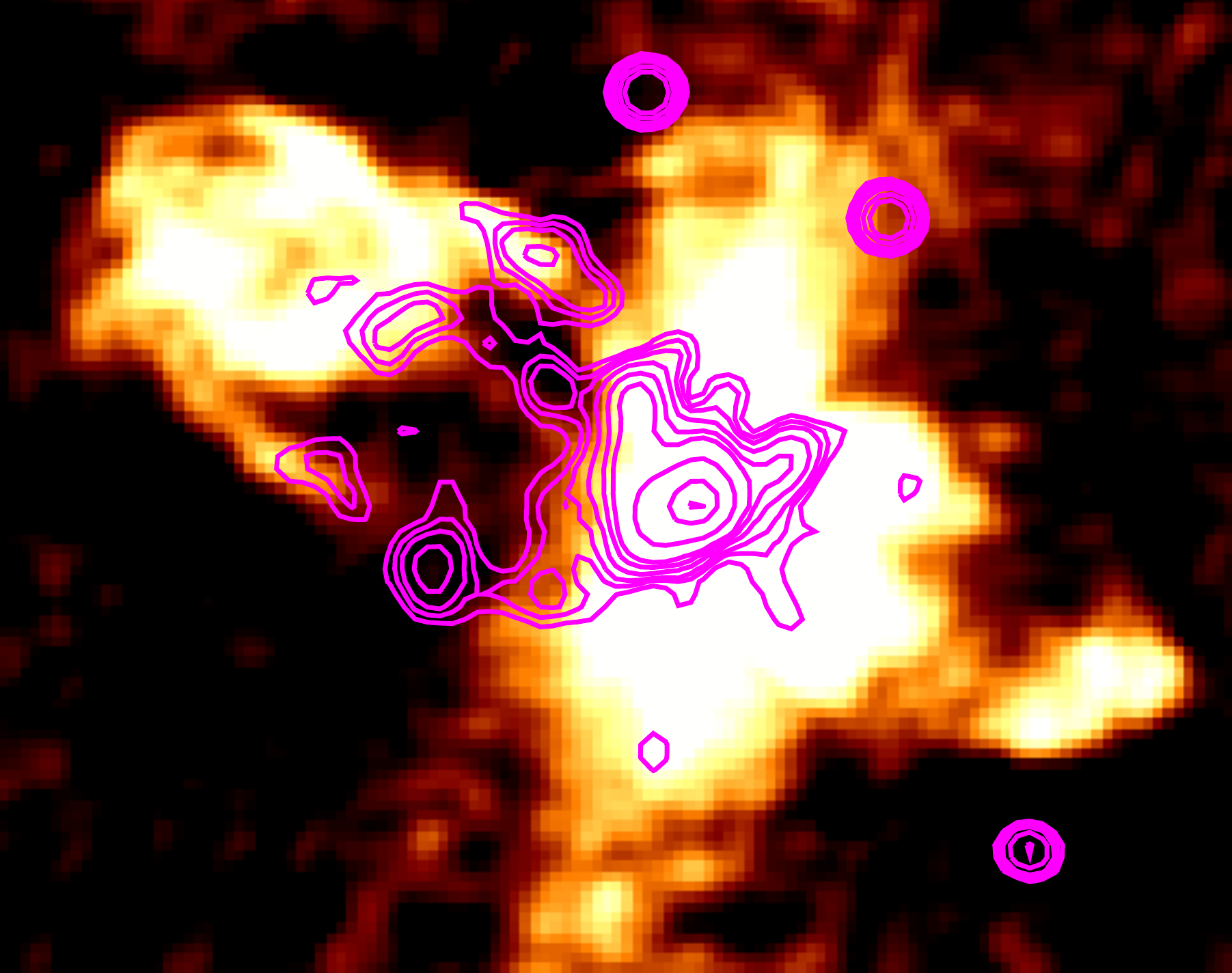}
}
\caption{ (Left) Overlay of the contours representing the {\it HST} WFPC2 $H\alpha +[NII]$ line emission in blue and (right) Chandra ACIS-S X-ray emission in pink, over the total intensity image in colour of NGC~3079 with the VLA A-array at 1.4~GHz.}
\label{fig7}
\end{figure*}

The three prominent radio filaments, (A, B and C) mentioned in Section~\ref{secmorph}, at first glance, appear to be aligned with the ionized filaments seen in the [N II] $\lambda6583 + $H$\alpha$ line emission image \citep{cecil2001}.

Figure~\ref{fig7} shows the overlay of the {\it Hubble Space Telescope} WFPC2 $H\alpha +[NII]$ emission line contours and Chandra ACIS-S X-ray emission contours over the radio continuum VLA A-array image at 1.4 GHz.
There is no one-to-one correspondence between the radio and the thermal emission filaments (see Figure~\ref{fig7}).

The [N II] $\lambda6583 +$ H $\alpha$ line emission image shows no counterpart for the south-western lobes due to extinction, although there is soft-X-ray emission \citep{cecil2002}. Among the three radio filaments that seemingly arise from the radio core, the filament marked C is aligned with the $H\alpha$ filaments. Such a correlation was previously seen in M82 by \citet{adebahr2013}. They suggested that this could be a proof for magnetic fields that are frozen into the ionized medium and get transported along with the superwind.

However, there are several other observational signatures in NGC\,3079 that cannot be explained using a simple frozen-in magnetic field.
Although partially aligned, it has to be noted that the weakest radio filament (A) corresponds to the brightest $H\alpha$ filament, which is also aligned almost along the same direction as that of the parsec-scale VLBA jet. This trend goes against the frozen-in magnetic field picture, where one would expect a correlation in the brightness of the radio and $H\alpha$ filaments. The brightest radio filament being offset from the direction of the VLBI jet also poses a problem for these filaments being powered by jets, unless the jet is changing its direction or undergoing precession. Moreover, the middle radio filament annotated B in Figure~\ref{fig7} does not appear to be associated with the central $H\alpha$ filament.
\cite{cecil2002} have shown that X-ray and $H\alpha$ filaments were spatially well correlated using their high-resolution images from Chandra and HST respectively. However, radio emission which is also mostly filamentary in nature does not seem to be well correlated with the thermal filaments. In the northern lobe, the closed loop like structure appears displaced northwards compared to the emission line filaments. The dearth of emission lines and X-ray emission at the topmost loop is intriguing. The radio emission also extends farther beyond the point where the emission line gas reaches a blowout phase as a result of the growth of Rayleigh-Taylor instabilities \citep{cecil2001}.

The radio plasma and thermal emission are not completely intermixed, which would imply different sources of origin. Such lack of intermixing despite the disruption of the jet at early stages is also witnessed in X-ray cavities in clusters of galaxies, where the ambient hot gas is simply pushed aside by the expanding radio source \citep{birzan2004}.
Similar to NGC\,3079, NGC\,253, a starburst galaxy hosting superwinds possesses magnetic fields which are aligned along the filaments. \cite{heesen2011} model the walls of the outflow cones of the galaxy to be threaded by helical magnetic fields as indicated by RM inversion. They appear as filaments in projection. But unlike NGC\,3079, there are only two filaments each at the edge of the outflow cone. If the inner $H\alpha$ or radio filaments actually trace the wall of a superbubble in the case of NGC\,3079, the expectations would be to see edge-brightened structures in projection or a completely filled volume rather than discrete filaments at intermediate angles.

The presence of separate filaments in between the edge filaments in NGC\,3079 casts doubts over the simple picture of these being the edge of the walls seen in projection. 
\cite{cecil2001} have studied the rotation measure of the northern ring of NGC\,3079. They have found that the RM changes its sign from the $+50~rad~m ^{-2}$ on the inside of the ring to $-50~rad~m^{-2}$ on the outside. We note that our EVLA in-band RM image exhibits average RM values in the same range and also shows the RM inversion. In our RM image of the northern lobe (see Figure~\ref{fig5}) we see the RM inversion in individual filaments on the right hand side although the polarized emission with positive RM has not yet been recovered. 


Below we consider plausible physical mechanisms that might be a playing a role in the formation of the ``ring"  in NGC\,3079. 


\subsection{Shock compression at the boundaries of an expanding superbubble}
\label{shocksec}
\cite{Duric88} propose that shocks at the interface between ambient medium and the outflow would be a natural consequence of the supersonic velocities derived for the winds present in NGC\,3079.
Several features seen in the ring of NGC\,3079 are suggestive of the presence of a shock. With our better resolution images, it can be seen that the ring-like structure is also comprised of several filaments. The entire structure is reminiscent of a supernova remnant shell. 
Shocks can produce synchrotron emission via diffusive shock acceleration (DSA) or Fermi first order acceleration \citep{blandford1987}. The power-law index of the energy distribution of the particles is determined by the strength of the shock, which can be described using the Mach number.

The flattening in the spectral index along the ridges of the ring at various locations might point towards the reacceleration of charged particles (see Section~\ref{spixpolprop}). The spectral index image shows that the spectrum is as flat as -0.5 at several locations which match with the mean spectral index values typically seen in galactic supernova remnants \citep{klein2018}.

By using the  spectral age of $\sim$2.8 Myr (see Table~\ref{tab3}) and an outflow length of 2.3 kpc, we obtain a  wind velocity of 800~km~s$^{-1}$. Since the velocity of the wind increases with the distance from the disk \citep{cecil2001}, our wind velocity represents an average value of the projected wind velocities at various distances above the disk. The thermal gas traced by emission lines exhibits radial velocity that increases with distance from the disk to attain velocities as high as 1000~km~s$^{-1}$ \citep{Veilleux1987,cecil2001}. Since most of the radio emission is from above the emission-line complex and assuming that the radio emission is coexisting with the superwind we can assume a radial velocity of 1000~km~s$^{-1}$ for the radio outflow as well. 
Hence, the total velocity of the radio  outflow is $v\sim1250$~km~s$^{-1}$.

The sound speed in the hot wind regions present in typical starbursts is about 100~km~s$^{-1}$ \citep{romero2018}. Hence the sonic Mach number turns out  to be $\sim$12. The spectral index, $\alpha$ depends on the Mach number of the shock \citep{blandford1987,vanweeren2016} as the following
\begin{equation}
\alpha=\frac{1}{2}-\frac{M^{2}+1}{M^{2}-1}
\end{equation}

For a sonic Mach number of 12, we obtain $\alpha \sim -0.5$. Hence the shock present in NGC\,3079 is strong enough to produce the spectral index values as flat as that seen in NGC\,3079 from an underlying thermal distribution of electrons.

It is also possible that an already existing relativistic population of electrons gets reaccelerated at the shock front. Detailed DSA simulations would be required to understand the contribution of the reacceleration of an already existing relativistic electron population.

Shock compression often leads to the ordering of the tangled or turbulent magnetic fields that permeate the ambient medium. The magnetic fields parallel to the shock front get amplified leading to high degrees of polarized emission aligned perpendicular to the shock fronts \citep{laing1980,ensslin1998}. The northern ring shows high fractional polarization compared to the rest of the galaxy. The polarization vectors are also aligned perpendicular to the individual filaments as expected in the case of shock amplification. The ``ring" also shows higher depolarization (see Section~\ref{spixpolprop}) which might be due to a higher Faraday depth within or in front of it \citep{cioffi1980}. So either the magnetic field or electron density (or both) is higher here.  It might be that the expansion of the shell (that we see projected as a ring) is sweeping up gas, so there is a denser shell of gas along the shell. It could also be that the magnetic fields are being compressed along the edges although compressing magnetic fields can lead to higher absolute polarization \citep{laing1980}.
 
Although the conditions in NGC\,3079 are favorable for the presence of shocks powerful enough to reproduce the observed spectrum, all the observed properties of NGC\,3079 cannot be explained using shock acceleration. The lack of correlation between the thermal and non-thermal emission suggests different mechanisms. More importantly, the RM inversion in filaments cannot be explained solely using shock acceleration, and requires organized magnetic fields. In the next section, we consider if the draping of magnetic fields at the leading edge of the outflow can address these shortcomings.

\subsection{Magnetic Draping}
\cite{bernikov1979,lyutikov2006} have pointed out that as an object moves through a weakly magnetized medium with a velocity greater than its Alfv\'en speed, it necessarily drags up a layer of magnetic field lines along with it. For magnetic draping to be playing a role it is required that the velocity of the wind is higher than the Alfv\'en velocity. \cite{adebahr2019} show that the radio lobes of B2\,0258+35 are probably being affected by magnetic draping using their polarization and HI study. We test this idea to explain the polarization structures observed in the lobes of NGC\,3079.

We estimate the Alfv\'en velocity of the wind using the following equation.
\begin{equation}
    v_{a}=2.18 n_{e}^{-\frac{1}{2}} B\,\,\,\,km\,s ^{-1}
\end{equation}
where B is the magnetic field in $\mu G$ and $n_e$ is the electron number density in cm$^{-3}$. For the northern lobes we have estimated an equipartition magnetic field  of $40\mu$G (see Table~\ref{tab3}). Our RM images show an average RM value of 50 rad m$^{-2}$. If we assume a Faraday screen path length of 100 parsecs, which is equal to the widths of the filaments as seen in our EVLA images, the electron number density turns  out to be 0.015 cm$^{-3}$ by using the relation RM =  $812 n_e B_{||} \ell$ rad m$^{-2}$. We obtain an Alfv\'en velocity  of $\sim700$~km~s$^{-1}$. Therefore, the wind is super Alfv\'enic with an Alfv\'enic Mach number, $M_{A}$ of 1.8. \cite{pfrommer2010} suggest that the following criteria have to be met for magnetic draping to occur,
\begin{equation}
\frac{\lambda_B}{R} \gtrsim \frac{1}{M_{A}}
\end{equation}
.0where $\lambda_B$ is the magnetic coherence length. The thickness of the polarized filaments in our image which is about 100 pc can be regarded as a lower limit of the magnetic coherence length. Assuming a radius of curvature, R $\sim$ 0.4 kpc, we obtain a value for the ratio of $\frac{R}{M_{A}}$  which is greater than the magnetic coherence length. Therefore, magnetic draping might not be a viable mechanism in the case of NGC\,3079.

\cite{guidetti2011} have suggested that RM inversion could be produced by magnetic field draping. However, it is not clear how any supersonic outflow which forms a draped magnetic field layer at its edge, can lead to a dome-like structure as observed in the northern lobe of NGC\,3079, rather than a conical structure. Hence, we suggest that  magnetic field draping does not play a significant role in the formation of the lobes in NGC\,3079.

\subsection{Filaments as magnetic flux tubes}
\label{magflux}

\cite{cecil2001} put forth three explanations for the observed RM inversion. In their first model, they consider that the wind entrains material from the cold disk with a frozen in magnetic field which is transported upstream and gives rise to the positive RM values. Once the disk material cools down the material starts falling back resulting in the reverse fields.

Secondly, they consider a globally ordered magnetic field of the halo, which surrounds the wind. The negative RM results from the compressed halo gas whereas the outflowing wind contributes to the positive RM.
In both these models, the outer shells lead to a positive RM while material inside leads to negative RM. However, as they had pointed out in Section 3.5 of their paper, Faraday rotation closely follows a $\lambda^2$ dependence which rules out a Faraday rotating medium within the synchrotron source. Therefore it is unlikely that the outflowing wind enclosed within either the host galaxy halo or the wind material that is falling back to the disk is giving rise to the positive RM.

They propose a third scenario which is similar to the solar prominence. Here, we consider the possibility of these being actual standalone filaments, in which case the RM inversion seen in the individual filaments can be easily explained using a helical magnetic field structure. The Faraday rotation is probably introduced by a sheath around the filaments with increased particle density and width about 10 \% of the filament size itself. The sheath which is threaded by a helical magnetic fields can introduce such an inversion. A helical magnetic field morphology is often invoked to explain the RM inversion seen in VLBI scale jets \citep{kharb2009,clausen2011,pudritz2012}.

\cite{henriksen2019} suggested that galaxies possess force-free magnetic fields that are generated by the $\alpha^2$ dynamo action; cosmic rays are transported in a direction parallel to the magnetic field lines under force-free conditions. Dome-like structures, as seen in NGC\,3079, and the active regions in the Sun, can be reproduced by some of the  solutions.

The morphology of radio filaments that are anchored to the core is especially reminiscent of the non-thermal radio filaments found in the Galactic center \citep{morris2014}. The Galactic center filaments also possess organized magnetic fields often aligned along the filament, similar to that seen in NGC\,3079. It is in principle possible that these filaments are magnetic flux tubes through which the relativistic material is transported from the core to the top.

Interestingly, it can also be inferred from Figure~\ref{fig1} that the middle filament, B is steeper compared to the edge filaments. The edge filaments might be tracing the shocked walls whereas the filament B might have a different source of origin and is completely unrelated to the other two filaments. This seems unlikely given the similarity in morphology. It is however possible that all the filaments are directed by the same dynamo mechanisms whereas the edge filaments are also being acted upon by shocks leading to a flatter spectrum. Hence, we conclude that this model successfully explains the structures and the RM inversion observed in NGC\,3079.

\section{Role of the jet in inflating the radio bubble}
\cite{irwinseaquist1988} argue that the large scale radio lobes and the VLBI scale jet are unusual in spiral galaxies and are therefore probably related. 
In this section, we investigate the role of the jet in actually inflating the radio bubbles.

Radio halos were discovered in several starburst galaxies \citep{ekerssancisi1977,allen1978,carilli1992,duric1998}. These were believed to be composed of the material that was convected up from the central starburst regions, where it was created by supernovae \citep{lerche1980}. However, these radio halos which are purely powered by starbursts have a very different morphology compared to NGC\,3079 \citep{Colbert96}. These radio halos are more spherical and diffuse whereas NGC\,3079 shows two distinctly directional lobes. 
The low-resolution images of NGC\,3079 show more extended and diffuse structures at lower frequencies (see Section~\ref{hostgal}) indicating a steep spectrum in the halo compared to the central `ring'. Hence, the nuclear lobes are distinct from the rest of the halo emission arising from the star-formation activity in NGC\,3079. 

Star formation rates (SFR) from the radio flux density of north-east lobes, south-west lobes and the disk are 2.58, 2.14 and 17.87 M$_\sun$~yr$^{-1}$ respectively \citep[using the relations in][]{condon2002} for stars more massive than $>$0.1~M$_\sun$. \cite{yamagishi2010} have estimated SFR using AKARI far-infrared data. They find that the SFR is 5.6 M$_\sun$~yr$^{-1}$ in total and 2.6 M$_\sun$~yr$^{-1}$ in the central 4 kpc region. The SFR estimated in the east and west lobes together exceed that estimated from IR emission in the central regions by about a factor of 2. Therefore, it is unlikely that the lobes are composed of plasma that is accelerated solely by supernovae and stellar winds. A certain contribution from the jet is inevitable unless the radio emission is generated as a result of in-situ particle acceleration in the lobes due to shocks (see Section~\ref{shocksec}). 

 
Energy and pressure requirements of the radio lobe can provide constraints on the origin of the bubble. \cite{Duric88} invoke accretion along with a nuclear starburst due to the very high rates of star formation required to meet the energy budget. Although the AGN at the core is weak, the radio jet on VLBI scales, as was pointed out by \cite{irwinseaquist1988}, is capable of powering the superbubble. 

\cite{irwinseaquist1988} compare the momentum flux that is required to feed the kpc-scale radio lobes to the pc-scale jets. They find that the momentum flux of the pc-scale jets is $3\times10^{31} h^{\frac{-10}{7}}\mathrm{g~cm~s}^{-2}$ while that of the north-eastern radio lobe is $1 \times 10^{32} h^{\frac{-10}{7}}\mathrm{g~cm~s}^{-2}$. However, these values are three orders in magnitude lower compared to the wind model parameters of \cite{Duric88}. Following \cite{irwinseaquist1988} we assume a slab volume with size 1.1 kpc. We find a momentum flux value of $1\times10^{33}h^{\frac{-10}{7}}\mathrm{g~cm~s}^{-2}$ for the north-eastern lobe for a filling factor value of 1.0. \cite{irwinseaquist1988} pointed out that the momentum flux of the lobes derived is a lower limit owing to the underlying minimum energy assumption. When we use a filling factor value of $1\times10^{-4} $ \citep{blustin2009} we find a momentum flux value of $3.5\times10^{35}h^{\frac{-10}{7}}\mathrm{g~cm~s}^{-2}$, which is comparable to the momentum flux derived from wind model parameters \citep{Duric88}. 
Interestingly, for the volume filling factor of $1\times10^{-4} $, the radio lobes are overpressured with respect to the surrounding hot gas \citep[e.g., see][]{Heckman90}. 

\cite{2015MNRAS.454.1404S} estimate the jet power of NGC\,3079 using various relations that connect radio luminosities to jet power. Their estimated value of 10$^{41}$ ergs s$^{-1}$ is two orders of magnitude lower than that required by the wind model in \cite{Duric88}. 
However the uncertainties in the relations used for conversion is often high that this difference is consistent with the wind model within error bars.

The radio continuum emission in NGC\,3079 is very different from the typical jet lobe morphology seen in powerful radio galaxies. Several VLBI studies had revealed that the jet in NGC\,3079 is expanding with mildly relativistic speeds in the range 0.1c to 0.18c \citep{trotter1998,sawada2000,kondratko2005}. It is also known that Seyfert galaxies often host very low power jets which often get disrupted or show S-shaped or twisted morphology \citep{ulvestad1981,booler1982,wilson1982,wilson1983,wehrle1988,Kharb06}.
Precession jet model can connect the VLBI jet to the large scale lobes with a jet precession period of $\sim 10^5$ yrs using the precession model of \cite{Hjellming81}. For instance, fixing the jet speed to be the one observed on VLBI scales ($v\sim0.2c$) yields a precession period of $9.5\times10^4$~yrs for a precessing cone half opening angle of $10\degr$ and jet inclination of $20\degr$, for the north-eastern lobe.  However, one episode of precession is insufficient to explain the entire complex morphology of the jet lobe structure.

\citet{henriksen1981} and \citet{fiedler1984} have invoked galactic pressure gradients and buoyancy to explain jet bending that produce S-shaped jets for certain viewing angles. \citet{smith1981} and \citet{wilson1982} have suggested that ram pressure bending due to a rotating ISM can produce S-shaped jets. However, this idea is difficult to test in the edge-on disk of  NGC\,3079, where the galaxy spiral arms are not clearly seen, making it difficult to ascertain the sense of the galactic ISM rotation. It is possible that the jet in NGC\,3079 got disrupted due to its interaction with the ambient medium or an accretion disk wind. 

This disrupted jet material probably gets transported out along with the superwind or along the force-free magnetic field lines generated via the dynamo mechanism,  or both. As discussed in Section~\ref{orig}, the lack of correlation of radio plasma with the superwind also suggests that the jet might be playing an important role. We conclude that the jet, the active nuclei, and the superwind together give rise to the peculiar lobe morphology in NGC\,3079 which is neither similar to a typical AGN jet or a starburst radio halo.

\section{Summary and Conclusions}
We have carried out a multi-frequency multi-scale radio study of NGC\,3079, a Seyfert 2 galaxy which also hosts a nuclear starburst. The origin of the radio continuum emission in this galaxy is widely debated. With our high resolution and high sensitivity data, we favor a scenario which involves an interplay of various processes and outflows. Our findings can be summarized as follows.
\begin{enumerate}
    \item There is no one-to-one correlation between the thermal emission (traced by X-ray and emission line) and non-thermal emission (traced by radio), which is inconsistent with magnetic fields frozen-in within the superwind hot material and being transported out.
    
    \item We propose that the filamentary structures are not merely edges of the superbubble seen in projection, but actual individual filaments (see Section~\ref{magflux}). The RM inversion observed in the individual filaments can also be easily explained using such a scenario. Each filament is like a jet segment threaded by helical magnetic fields. Structures similar to those seen in NGC\,3079 have been reproduced using dynamo models in the literature.
    
    \item  We find that the conditions in NGC\,3079 are favorable for shock acceleration to occur. However, it is insufficient to explain all the observed signatures, like RM inversion. Detailed simulations would be required to disentangle the role played by  shock acceleration in the resulting morphology.
    
    \item  The SFR estimated from the lobes suggest that supernovae and stellar outflows cannot feed the lobe material entirely and a contribution from the AGN jet is required. As discussed in section~\ref{shocksec}, shock acceleration of the thermal pool of electrons in the wind has not been ruled out. It is not entirely clear if all the emission can be accounted for by this mechanism. An observational test would be to check whether such a radio excess is seen in other starburst galaxies hosting superwinds in which an AGN core is absent.
\end{enumerate}

We conclude that the AGN jet contributes to the relativistic plasma that is observed in the lobes of NGC\,3079.  
We, therefore, propose that the radio lobes in NGC\,3079 are AGN jet-related. However, the presence of starburst superwind is also indicated, and it is hard to discriminate the effects of each mechanism. 

\section{Acknowledgements}
We thank the anonymous referee for the insightful suggestions, which led to the improvement of the manuscript. The National Radio Astronomy Observatory is a facility of the National Science Foundation operated under cooperative agreement by Associated Universities, Inc. This research has made use of the NASA/IPAC Extragalactic Database (NED) which is operated by the Jet Propulsion Laboratory, California Institute of Technology, under contract with the National Aeronautics and Space Administration.
S. Baum  and C. O'Dea are grateful to the Natural Sciences and Engineering Research Council of Canada (NSERC) for support.

\end{document}